\documentclass{ieeeaccess}
\usepackage{amsmath,amssymb,amsfonts}
\usepackage{algorithmic}
\usepackage{textcomp}
\usepackage{enumerate}
\usepackage{graphicx}
\usepackage[outdir=./pics]{epstopdf}
\usepackage{float}
\usepackage{lipsum,multicol}
\usepackage{subcaption} 
\usepackage{diagbox}
\usepackage{cite} 
\def\BibTeX{{\rm B\kern-.05em{\sc i\kern-.025em b}\kern-.08em
    T\kern-.1667em\lower.7ex\hbox{E}\kern-.125emX}}
\begin{document}
\history{Date of publication xxxx 00, 0000, date of current version xxxx 00, 0000.}
\doi{10.1109/ACCESS.2017.DOI}

\title{A Channel-Aware Routing Protocol With Nearest Neighbor Regression For Underwater Sensor Networks}
\author{\uppercase{Boyu Diao}\authorrefmark{1}, \IEEEmembership{Member, IEEE},
\uppercase{Chao Li\authorrefmark{1}, Qi Wang\authorrefmark{1}, Zhulin An\authorrefmark{1}, and Yongjun Xu\authorrefmark{1}
}}
\address[1]{Institute of Computing Technology, Chinese Academy of Sciences, Beijing (e-mail: diaoboyu2012@ict.ac.cn, xyj@ict.ac.cn, lichao@ict.ac.cn, wangqi08@ict.ac.cn, anzhulin@ict.ac.cn)}

\tfootnote{This paper is supported in part by National Natural Science Foundation of China (NSFC) under grant No.(61602447) and No.(61702487). The authors would like to thank all the people involved in KW14 and KAM11 sea trials.}

\markboth
{Boyu Diao \headeretal: A Channel-Aware Routing Protocol With Nearest Neighbor Regression For Underwater Sensor Networks}
{Boyu Diao \headeretal: A Channel-Aware Routing Protocol With Nearest Neighbor Regression For Underwater Sensor Networks}

\corresp{Corresponding author: Boyu Diao(e-mail: diaoboyu2012@ict.ac.cn).}

\begin{abstract}
The underwater acoustic channel is one of the most challenging communication channels. Due to periodical tidal and daily climatic variation, underwater noise is periodically fluctuating, which result in the periodical changing of acoustic channel quality in long-term. Also, time-variant channel quality leads to routing failure. Routing protocols with acoustic channel estimation, namely underwater channel-aware routing protocols are recently proposed to maintain the routing performance. However,  channel estimation algorithms for these routing protocols are mostly linear and rarely consider periodicity of acoustic channels. In this paper, we introduce acoustic channel estimation based on nearest neighbor regression for underwater acoustic networks. We extend nearest neighbor regression for SNR (Signal-to-Noise Ratio) time series prediction,  providing an outstanding prediction accuracy for intricately periodical and fluctuating received SNR time series. Moreover, we propose a quick search algorithm and  use statistical storage compression to optimize the time and space complexity of the algorithm. In contrast with linear methods, this algorithm significantly improves channel prediction accuracy (over three times at most) on both simulation and sea trial data sets. With this channel estimation method, we then propose a Depth-Based Channel-Aware Routing protocol (DBCAR). Taking advantage of depth-greedy forwarding and channel-aware reliable communication, DBCAR has an outstanding network performance on packet delivery ratio, average energy consumption and average transmission delay which is validated through extensive simulations.
\end{abstract}


\titlepgskip=-15pt

\maketitle

\section{Introduction}
{D}{uring} the past decade, there has been a significant development in underwater acoustic communication, thus inspiring extensive research on underwater sensor networks (UWSNs). At the same time, marine applications on underwater sensor networks, including ocean exploration, submarine tracking, and marine rescue \cite{cui2006challenges, practical_issues2007, hgj_secure_communication} have gradually matured. Long-term and reliable data transmission is the primary research object for UWSNs which faces many challenges: long transmission delay as the propagation speed of acoustic waves in water is approximately 1500 m/s, which is five orders of magnitude lower than that of radio frequency\cite{speed1500_kam2011impact}; limited-bandwidth as it is difficult for the bit rate of acoustic channels to exceed 100 kbps in a long-range system \cite{6757189}; time variable channel as underwater acoustic channel affected by ambient noise, leading to severe bit error\cite{cui2006challenges}.


Due to periodical tidal and daily climatic variation, underwater noise is periodically changing, result in acoustic channel quality changing with a periodical trend in long-term.   Since underwater acoustic channels are time variable, it is useful to estimate channel anterior to data transmission, in order to explore reliable communication links. Also, time-variant channel quality leads to routing failure. Routing with acoustic channel estimation, namely underwater channel-aware routing protocols are recently proposed to maintain the performance of routing protocols\cite{RN96, marlin, 6964460, RN213}. These protocols are able to investigate a reliable communication path from sources to sinks hop by hop. In channel-aware routing protocols, acoustic channel quality estimation is the core process of the best forwarder selection. 
With a group of quality indicators,  channel quality estimation can be modeled as a time series analysis problem. Simple mean statistical analysis, exponential moving average (EMA) and auto-regression (AR) were adopted for underwater channel quality series analysis. However,  these algorithms are linear and rarely consider periodicity of acoustic channels. With periodical tidal and daily climatic variation, underwater noise is changing periodically in long-term\cite{Nathan2016Underwater}. Meanwhile, the fluctuations of SNR values exist intrinsic periodicity in long-term underwater communication. 

In this paper, we consider intrinsic periodicity in long-term underwater acoustic channels and propose the nearest neighbor regression (NNR) based channel estimation algorithm. Moreover, we propose a depth-based channel-aware routing (DBCAR) protocol with NNR channel estimation algorithm. Depth-based routing protocols\cite{dbr2008, Diao2017A, Diao2015Improving} are location-free and stateless which have unstable packet delivery ratios in sparse networks but low transmission delay. On the contrary, channel-aware routing protocols have better packet delivery rates but high transmission delay. DBCAR with NNR channel estimation provides a more powerful solution with excellent packet delivery ratios while keeping a reasonable transmission delay. In summary, we make two significant contributions for underwater acoustic channel estimation and channel-aware routing protocols: 1) We introduce NNR for underwater channel quality estimation, and extend NNR for time series analysis, thus providing a better prediction precise for intricately periodical and fluctuating received SNR time series. In addition, we use hash table and statistical storage compression to optimize the time and space complexity of NNR quality estimation algorithm. 2) Based on the NNR quality estimation algorithm, we propose a depth-based channel-aware routing protocol, DBCAR. Besides residual energy, we take historical statistical parameters, current SNR gradient, and depth into consideration to explore the best forwarder hop by hop. The accuracy of NNR quality estimation algorithm has been evaluated through simulations on different SNR fluctuation scenarios and sea trials data sets, including ideal periodic fluctuation, periodic fluctuation with stochastic noises and random periodic fluctuation, and underwater experiment data, KW14 and KAM11. The performance of DBCAR has been evaluated through extensive simulations using Bellhop\cite{Bellhop} and Aqua-sim\cite{aqua2009xie}.


The rest of the paper is organized as follows. Selected works on acoustic channel estimation methods and channel-aware routing protocols are summarized in Section \ref{related work}. Proposed channel estimation algorithm and a channel-aware routing protocol are presented in Section \ref{protocol description}, including NNR channel quality estimation algorithm, time and space complexity optimization and DBCAR, a depth-based channel-aware routing protocol with NNR. Extensive evaluations and results are analyzed in Section \ref{simulation}. Section \ref{conclusion} concludes this paper.

\section{Related Work}
\label{related work}

Acoustic channel quality is time-variant \cite{820739}  while most underwater network protocols assumed acoustic channel quality is static \cite{vbf2006, dbr2008, Han2015Routing}, leading to unsatisfying network performance, including low packet delivery ratio, low throughput, and redundant re-transmissions in real underwater environments. Channel estimation algorithms considering time variability of acoustic propagation results in a better channel estimation accuracy. Besides, considering time-variant channel quality provides a practical performance evaluation for network protocols\cite{Tomasi2013Impact}. Underwater routing protocols with channel quality estimation, namely channel-aware routing protocols\cite{RN96,marlin,6964460,RN213} can investigate a reliable communication path from sources to sinks hop by hop, maintaining a remarkable network performance in time-variant underwater acoustic channel states. 


Acoustic channel quality estimation is the core process for channel-aware routing protocols. Historical and current packet success ratio\cite{RN96,marlin} and received signal-to-noise ratio \cite{6964460,RN213} were taken into account as the primary channel quality indicator for estimation and prediction. Packet success ratio (PSR) is the ratio of successful delivery packets and total transmitted packets counted hop-by-hop between each node and its neighbors. Received signal-to-noise ratio (SNR) is computed as Equation \ref{snr} where $P$ is the transmission power, $A(r,f)$ is the attenuation over a distance of r with a signal of frequency f. $N(f)$ is the noise power spectral density and $\Delta f$ is the receiver noise bandwidth\cite{snr2006}. Packet success ratio and received signal-to-noise ratio have an equivalence relation from Equation \ref{snr and ratio}\cite{4752682}. Besides packet success ratio and received signal-to-noise, hops away from sinks, residual energy, and other auxiliary parameters were also considered for channel quality estimation and best forwarder selection.

\begin{equation}
SNR=\frac{P/A(r,f)}{N(f) \Delta f}
\label{snr}
\end{equation}

\begin{equation}
PSR=1 - (1 - \frac{1}{2} erfc(\sqrt{SNR})) ^ L
\label{snr and ratio}
\end{equation}

%

\begin{figure}[H]
\centering
\includegraphics[width=3in]{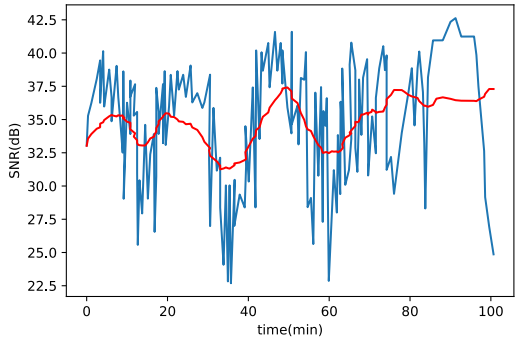}
\caption{SNR samples in KW14 sea trial }
\label{kw14}
\end{figure}

\begin{figure}[H]
\centering
\includegraphics[width=3in]{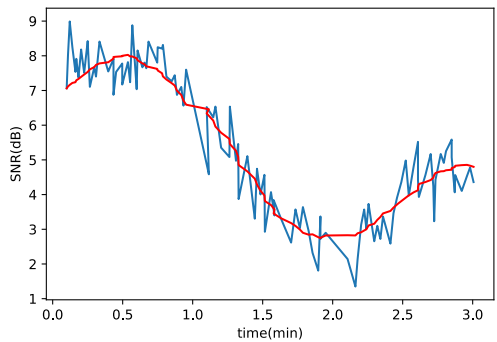}
\caption{SNR samples in KAM11 sea trial }
\label{kam11}
\end{figure}

Channel quality estimation can be modeled as a time series analysis process. Simple mean statistical analysis\cite{RN213}, exponential moving average (EMA)\cite{RN96, RN217, marlin} and auto-regression (AR)\cite{RN196} were adopted for underwater channel quality time series analysis. The EMA for a series $Y$ may be calculated recursively as Equation \ref{EMA}. For AR, the notation AR(p) indicates an auto-regressive model of order p defined in Equation \ref{AR}. However, linear time series analysis methods are not precise for underwater link quality estimation as SNR intricately fluctuating with ambient noise. Also, SNR fluctuation exists intrinsic periodicity in long-term underwater communication\cite{Nathan2016Underwater}. Also, these conclusions can be drawn from real ocean experiments, including KW14 (Fig.\ref{kw14}) and KAM11(Fig.\ref{kam11}).  KW14 sea trials took place August 2014 in Keweenaw Waterway near Michigan Tech. Received SNR time series in decibels are presented in Fig.\ref{kw14}, the receiver is 312m away from the transmitter. Kauai Acomms MURI 2011 (KAM11) Experiment took place from June to July 2011 in waters around the coast of Martha's Vineyard (MA, USA), Pianosa (Italy), and Kauai (HI, USA) islands, respectively. SNR measurements taken on July 8 were shown in Fig.\ref{kam11} \cite{kam11, Tomasi2013Impact}.

\begin{eqnarray}
\label{EMA}
S_t =
\begin{cases}
Y_1, & t = 1 \cr  \alpha \cdot Y_t +(1 - \alpha) \cdot S_{t-1}, & t > 1
\end{cases}
\end{eqnarray}

\begin{equation}
\label{AR}
X_t = c + \sum ^ p _{i=1} \phi _i X_{t-i} + \epsilon_t
\end{equation}

With acoustic channel quality estimation, some channel-aware routing protocols were proposed \cite{RN96, marlin, RN213, RN217, RN254}. In \cite{RN217}, a cross-layer routing protocol for UWSN was proposed which exploited link quality information for cross-layer relay determination. In addition, simple network topology parameters, including hop count, were considered to improve routing performance. EMA was selected as the major channel estimation method. \cite{RN96, marlin} extended ideas in \cite{RN217} to sea trials and multi-modal communications. In \cite{RN213} an energy efficiency channel-aware and depth-based routing protocol was proposed with autonomous underwater vehicles. Besides channel quality, this paper considered node energy, hop count and propagation delay to improve routing performance. Average peak signal-to-noise vectors (PSNR)\cite{psnr} were used to estimate channel quality. In \cite{RN254}, a channel-aware, depth-adaptive routing protocol was proposed, taking sound speed and underwater noises into account to relief the rate of transmission error. On relay selection, three factors were taken into consideration: successful transmission probability, the distance between candidate nodes and the destination, and the distance between candidate nodes and the ideal path. The successful transmission probability was computed between a source node and its neighbors respectively.

 \begin{figure}[H]
\centering
\includegraphics[width=3in]{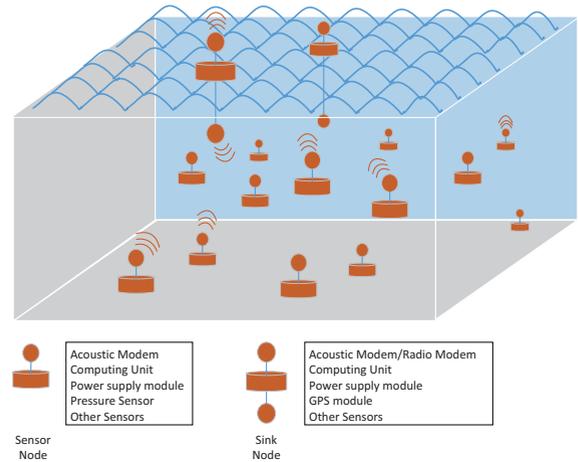}
\caption{Network Architecture}
\label{network_architecture}
\end{figure}

Different from above algorithms and protocols, we introduce NNR for underwater channel quality estimation in this paper, extending NNR for time series analysis and providing a better prediction precise for intricately periodical and fluctuating received SNR time series. However, NNR based channel estimation algorithm is non-linear, thus increasing time and space complexity than linear algorithms. Therefore, we use hash table and statistical storage compression to optimize the time and space complexity of NNR quality estimation algorithm, getting an approximate linear performance. Based on the NNR quality estimation algorithm, we proposed a depth-based channel-aware routing protocol (DBCAR). Besides residual energy, we take historical statistical arguments, current SNR gradient, and depth into consideration to explore the best forwarder hop-by-hop. We will present proposed algorithm and protocol in the next section.

\section{Algorithm and Protocol Description}
\label{protocol description}

In this section, we start by describing the network model before defining the initialization process. Then, we present NNR channel quality estimation algorithm together with time and space complexity optimization methods. Finally, we present a depth-based channel-aware routing protocol.

\subsection{Network Model}
\label{network_model}

We constitute the network within a cube area. $N$ underwater sensor nodes are vertically deployed in the area and slowly moving with water currents. As shown in Fig.\ref{network_architecture}, sensor nodes are uniformly deployed. Besides, there is one or multiple sinks on the surface. Each sensor node has a unique identification number and the same transmission power. Channel attenuation function is defined in Equation \ref{attenuation function}.

\begin{equation}
\label{attenuation function}
A(l, f)=l^p a(f)^l
\end{equation}

\begin{equation}
\label{attenuation function dB}
10 log A(l,f) = p \cdot 10 log l + l \cdot  10 log a(f)
\end{equation}

where $l$ is transmission distance, $f$ is carrier frequency. $A(l,f)$ is acoustic channel attenuation function in units of $dB$ given by Equation \ref{attenuation function dB}, $p \in [1,2]$, $a(f)$ is the absorption coefficient. 


Once an underwater node deployed into the water and switched on, it will broadcast a handshake packet to explore neighbors and initialize a vector of SNR vectors, notated as $M_{SNR}$ based on ACK packets from neighbors given by Equation \ref{channel list}, where $X^{ID_1}$ is the received SNR vector of neighbor $ID_1$. For the NNR channel quality estimation algorithm, the input vector is an element of $M_{SNR}$, notated as Equation \ref{channel vector}, where $x_{t_1}$ notates SNR values is X from a neighbor with $ID_i$ at $t_1$ moment.

\begin{equation}
\label{channel list}
M_{SNR} = \left[
 \begin{matrix}
 X^{ID_1} & X^{ID_2} & \cdots & X^{ID_i}  & \cdots & X^{ID_n}
  \end{matrix}
  \right] 
\end{equation}

\begin{equation}
\label{channel vector}
 X^{ID_i}=\left(
 \begin{matrix}
   x_{t_1} & x_{t_2} & x_{t_3} & \cdots\  & x_{t_m}
  \end{matrix}
  \right)
\end{equation}

\subsection{NNR Channel Estimation Algorithm}
\label{NNR Channel Estimation Algorithm}

In this section, we extend NNR to the time series prediction problem. In the primal NNR algorithm, the training set contains a group of vectors taking values in $\mathbb{R}^h$ and $y$ is the output label of each vector given in Equation \ref{primal NNR}. 

\begin{equation}
\label{primal NNR}
 \begin{matrix}
   (x_1^1 & x_2^1 & \cdots & x_h^1) & \sim  & y_1 \\
   (x_1^2 & x_2^2 & \cdots & x_h^2) & \sim  & y_2 \\
   (x_1^3 & x_2^3 & \cdots & x_h^3) & \sim  & y_3 \\
   \vdots & \vdots & \vdots & \vdots \\
   (x_1^n & x_2^n & \cdots & x_h^n) & \sim  & y_n \\
  \end{matrix}
\end{equation}

We want to predict the output label $y_{n+1}$ of $(x_1^{n+1}, x_2^{n+1}, \cdots, x_h^{n+1})$ based on the training set. $k$ nearest neighbor vectors are taken from the training set by distance function (e.g., Euclidean distance function given by Equation \ref{distance function}). 

\begin{equation}
\label{distance function}
D(X^1, X^2) = \sum^n_{i=1}(x_{i}^1 - x_{i}^2)^2
\end{equation}

Then, output label $y$ of the input vector is determined by Equation \ref{y function} with inverse distance weighting of the $k$ nearest neighbors where $q$ is a negative integer, and $d_i$ is the distance. Inverse distance weighing is a weighted average method that weight decreases as distance increases when computing weighted average. 

\begin{equation}
\label{y function}
y = \sum^{k}_i \omega_i y_i    \qquad  \omega_i = \frac{d_i^q}{\sum^k_i d_i^q}
\end{equation}

However, in a time series prediction problem, there is no direct output label $y$ but a vector of values with the same property and constantly accumulate with time, like the time series in Expression.\ref{time extend}. 

\begin{equation}
\label{time extend}
 \begin{matrix}
   x_{t_1} & x_{t_2} & x_{t_3} & \cdots\  & x_{t_{n-m}} & \underbrace{x_{t_{n-m+1}} \quad  x_{t_{n-m+2}}  \cdots  x_{t_{n}}}_{m}
  \end{matrix}
\end{equation}

To predict $y_{t+1}$, we define a time window with a length of $m$. We use the latest $m$ SNR values as an input vector of NNR, then sliding the time window from the beginning of the SNR vector, setting $m+1$ SNR value as corresponding output $y$, thus getting a training set notated in Expression.\ref{extension} ($m = 3$). With the training set, we can calculate distances step by step, getting $k$ nearest vectors in $m$ dimension. With $k$ nearest vectors, we get $k$ prediction values, constituting the prediction vector. For example, we assume $k=2, m=3, n=10$, and we want to predict $x_{11}$. Therefore, the input vector of NNR is $(x_8, x_9, x_{10})$. Then, we assume the 2 nearest neighbor is $(x_1, x_2, x_3)$ and $(x_4, x_5, x_6)$ with distances $d_1$ and $d_4$ away from the input vector. Thus, we get the prediction vector $(x_4, x_7)$. With inverse distance weighting function given by Equation \ref{y function}, we get prediction value of time $x_{11}=\frac{d_1^p}{d_1^p+d_4^p} x_4+\frac{d_4^p}{d_1^p+d_4^p} x_7$ eventually, where $p$ is generally set to be $2$.

\begin{equation}
\label{extension}
 \begin{matrix}
   (x_1 & x_2  & x_3) & \sim  & x_4 \\
   (x_2 & x_3  & x_4) & \sim  & x_5 \\
   (x_3 & x_4  & x_5) & \sim  & x_6 \\
   \vdots & \vdots & \vdots & \vdots \\
   (x_{n-3} &  x_{n-2}  & x_{n-1}) & \sim  & x_n 
  \end{matrix}
\end{equation}

The time complexity of this algorithm is $O(nm)$ and space complexity is $O(n)$ where $n$ is the length of an SNR vector, and $m$ is the length of sliding window. However, it is not an efficient solution so that we will optimize the time and space complexity of this algorithm in the next section.

\subsection{Algorithm Optimization}
\label{optimization}

\noindent \textbf{Time Complexity Optimization:} We adopt a hash table to store SNR records. Multiplying each SNR by $10^3$ and keeping the integral part to form a new SNR vector. Using an arbitrary hash function to compute an index. We get hash keys from SNR values and hash values are the corresponding time stamps of SNR values, then putting SNR values into an array of buckets, from which the time stamp of an expected SNR value can be directly found. When finding $k$ nearest neighbors, we shorten searching range after the $min$ values upgraded. In general, we usually initial the $k$ $min$ values with a big number, e.g. 65535. Therefore, the initial searching interval is the whole SNR vector. During the calculation, we update the $k$ $min$ values while we get a distance smaller than the max value of $k$ $min$ values. Once the $k$ $min$ values updated, we shorten the search range with recently updated $min$ value. The searching interval is given in Expression \ref{search range} where $ x_{t_{n-m+1}}$ is defined in Expression \ref{channel vector}.

\begin{equation}
\label{search range}
\left [ \left \lfloor x_{t_{n-m+1}} - \sqrt{min} \right \rfloor, \left \lceil x_{t_{n-m+1}} + \sqrt{min} \right \rceil \right ] 
\end{equation}

\noindent \textbf{Space Complexity Optimization:}   Length of an SNR vector is continuously increasing with time, consuming lots of hard disk space. SNR values far from now have little significance for current prediction. So we employ time inverse distance weighting like Equation \ref{compression function} to compress prior SNR values. We set a storage limit $L$ and a period $T$. When the length of SNR vector reaches $L$ or the sensor node operates for $T$ hours, the compression process will be triggered. The first $\alpha L$ SNR values will be compressed into an inverse distance weighting average, then the length of SNR vector is $(1-\alpha)L$, where factor $\alpha \in [0, 0.5]$.

\begin{equation}
\label{compression function}
y = \sum^{\alpha L}_i \omega_i x_i    \qquad  \omega_i = \frac{d_i^p}{\sum^L_i d_i^p}
\end{equation}


Then we will prove that the time complexity of the optimized algorithm will be reduced by orders of magnitude. Underwater channel quality is mainly affected by marine environmental parameters such as hydro-meteorology. The statistical law of general meteorological data conforms to the normal distribution. Therefore, we assume that the channel quality satisfies the discrete normal distribution in the proof of algorithm complexity\cite{KEMP1997223}. According to the conclusions of Szablowski\cite{SZABLOWSKI2001289}, while the variance $\sigma ^2 > 0.73$, the probability density of discrete normal distribution can be approximated by Equation \ref{discrete_normal} where $\alpha$ is the mean and $\sigma$ is the variance.

\begin{equation}
\label{discrete_normal}
P(X=i) = \frac{1}{\sigma \sqrt{2\pi}} e^{-\frac{(i-\alpha)}{2\sigma^2}}
\end{equation}

\begin{figure}
   \centering
    \includegraphics[width=9.5cm]{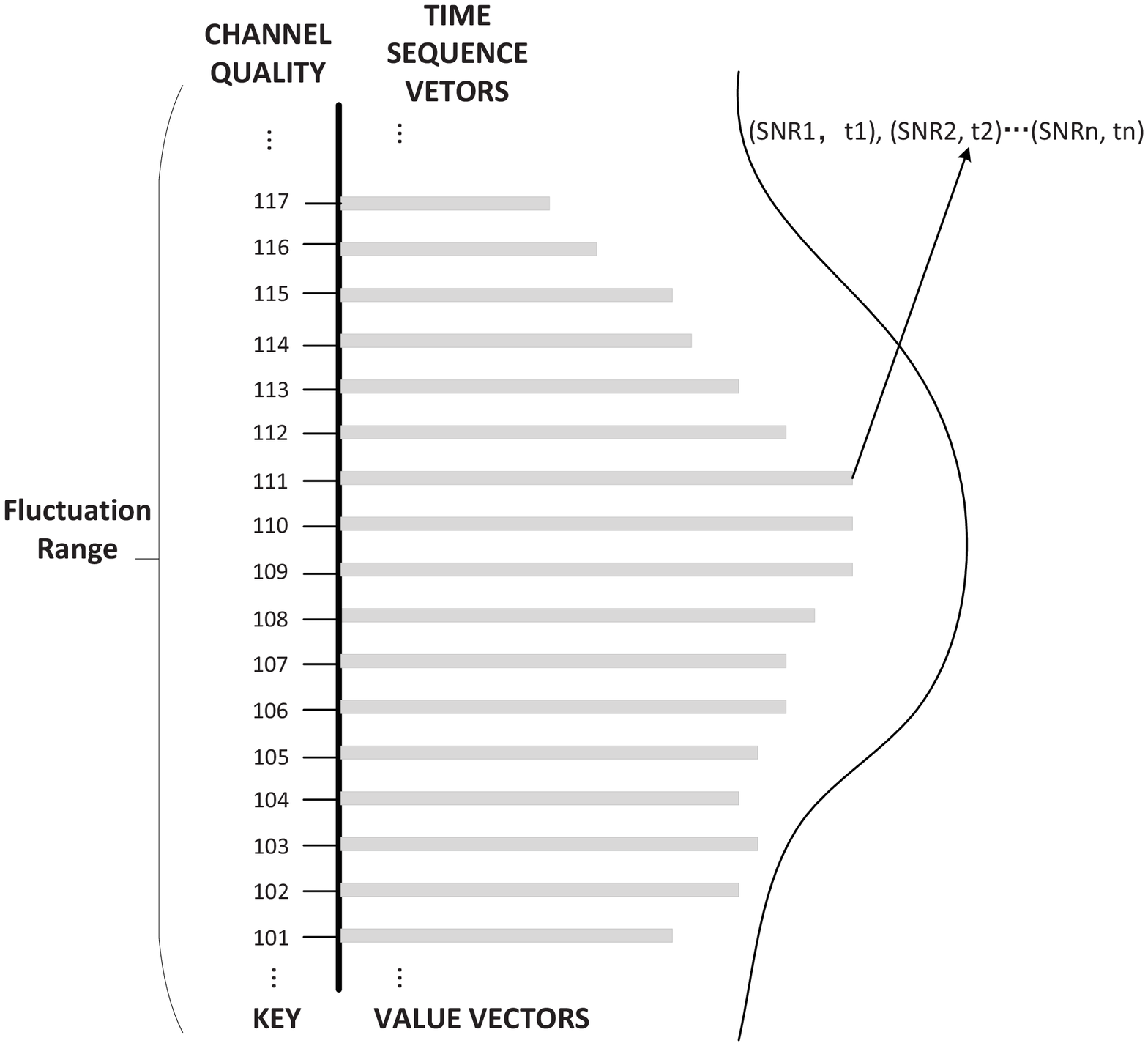}
    \caption{SNR Hash Table}
    \label{fig:hash_snr}
\end{figure}

Setting the channel quality values fluctuate within $r=|v_{max} - v_{min}|$, keeping the channel quality two decimal before hashing, that is the same with multiplying all channel quality values by 100 and retain the integer part. And now the values fluctuate within $100r$, so the total number of keys in the hash table in Fig. \ref{fig:hash_snr} is at least $100r$


Total number of nearest neighbor calculations is related to the length of timing sequence vectors $n$, the length of the time window for calculating the nearest neighbor regression $m$ and the value of target channel quality $i$, denoted as $T(X=i,m,n)$. 
When there are $n$ keys, the expected length of the value vector that is randomly selected in the hash table is $n\frac{1}{\sigma \sqrt{2\pi}} e^{-\frac{(s-\alpha)}{2\sigma^2}}$. According to proposed optimized algorithm, $T(X=i,m,n)$ is the sum of the $m$ value vector lengths of all the data in the hash table (Equation \ref{discrete_normal_proof1}).
 
\begin{equation}
\label{discrete_normal_proof1}
\begin{aligned}
T(X=i,m,n) &= \frac{n}{\sigma \sqrt{2\pi}}(\frac{1}{e^{\frac{i_1-\alpha}{2\sigma^2}}} + \frac{1}{e^{\frac{i_2-\alpha}{2\sigma^2}}} + ... \frac{1}{e^{\frac{i_m-\alpha}{2\sigma^2}}}) \\
&\leq \frac{nm}{\sigma \sqrt{2\pi}}(\frac{1}{e^{\frac{i-\alpha}{2\sigma^2}}})\\
\end{aligned}
\end{equation}


When $r\geq 10$, we have $100r \geq 1000$. The distance between randomly selected key $i$ and the expectation key $\alpha$ is denoted as $E(i-\alpha)$. Then we have Equation \ref{discrete_normal_proof2}.

\begin{equation}
\label{discrete_normal_proof2}
\begin{aligned}
E(i-\alpha) &\geq \frac{\sum\limits_{i=1}^{500} i}{500}  \geq \frac{500(500+1)}{1000} \\
&\geq 250
\end{aligned}
\end{equation}


While $1 < \sigma ^ 2 < 12.5$, we have $\sigma\sqrt{2\pi} e^{\frac{i-\alpha}{2\sigma^s}} \geq \sigma\sqrt{2\pi}e^10 \geq e^{11}$. Then Equation \ref{discrete_normal_proof1} can be further organized as Equation \ref{discrete_normal_proof3}.

\begin{equation}
\label{discrete_normal_proof3}
\begin{aligned}
T(X=i,m,n) &= \frac{n}{\sigma \sqrt{2\pi}}(\frac{1}{e^{\frac{i_1-\alpha}{2\sigma^2}}} + \frac{1}{e^{\frac{i_2-\alpha}{2\sigma^2}}} + ... \frac{1}{\frac{i_m-\alpha}{2\sigma^2}}) \\
&\leq \frac{nm}{\sigma \sqrt{2\pi}}(\frac{1}{e^{\frac{i -\alpha}{2\sigma^2}}})\\
&\leq \frac{nm}{\sigma \sqrt{2\pi}e^{10}}\\
&\leq \frac{nm}{e^{11.5}}  \leq O(\frac{nm}{10^{4.8}}) 
\end{aligned}
\end{equation}

Finally, we get that, when $r>10, 1 < \sigma ^ 2 < 12.5$, proposed optimized algorithm is able to reduce time complexity by approximately 5 orders of magnitude.


\subsection{DBCAR: a depth-based channel-aware routing protocol}

In this section, we apply NNR-based channel estimation algorithm to depth-based routing protocols, then propose a depth-based channel-aware routing protocol, DBCAR. Depth-based routing protocols are location-free and stateless with unstable packet delivery rates in sparse networks but low transmission delay. On the contrary, channel-aware routing protocols have better packet delivery rates but high transmission delay. DBCAR with NNR channel estimation provides a more powerful solution among depth-based routing protocols and channel-aware routing protocols with outstanding packet delivery rates while keeping a reasonable transmission delay. 

\begin{figure}[H]
\centering
\includegraphics[width=3in]{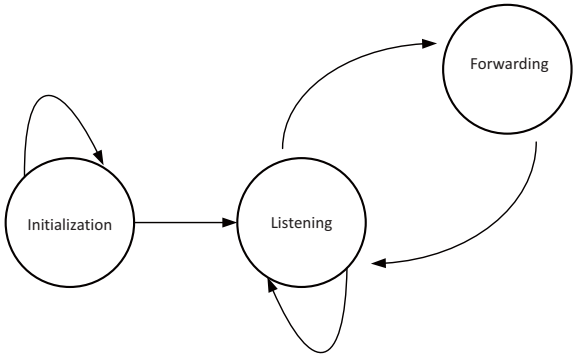}
\caption{Transition Relationship between Operation Processes}
\label{three_process}
\end{figure}

In DBCAR, each underwater node has three operation processes, including the initialization process, the listening process, and the forwarding process. Transition relationship between the three processes is described in Fig.\ref{three_process}. 

Each underwater node starts initialization process once deployed into the water and switched on. In the initialization process, each node broadcasts a handshake packet containing node ID and initial energy volume. Every neighbor node that receives the handshake packet will reply an ACK packet containing node ID and residual energy volume. With ACK packets and the corresponding received SNR values, nodes in initialization process construct the list of SNR vectors described in Section \ref{network_model}. Besides the list of SNR vectors, we utilize residual energy which can be obtained in a variety of ways\cite{kim2015simple}\cite{mishra2015residual}, historical SNR means, and variances as auxiliary parameters for best forwarder selection. After the initialization process, each underwater node has a list of neighbor nodes information, including received SNR vectors, latest residual energy and historical SNR means and variances, then changes to the listening process. 

In the listening processes, each node keeps listening and analyzing every received packets. When parsing received packet, the node obtains received SNR, residual energy, forwarder ID and other information of the source node. We assume a node $A$ in the listening process receives a packet from node $B$. Then node $A$ will update received SNR vectors and other parameters corresponding to node $B$ in $M_{SNR}$ (defined in Equation \ref{channel vector}). If the forwarder ID is the same with node $A$, this node will change to forwarding process. Because node $B$ has chosen node $A$ as the best forwarder. Otherwise, node $A$ maintains in the listening process. To calculating mean and variance of received SNR values based on time series, we introduce a recursive algorithm utilizing only the latest arrived value, latest mean and variance. This algorithm has a time complexity of $O(n)$ in contrary to the naive calculation of mean and variance using all historical data with the time complexity of $O(n^2)$. The mean and variance calculation methods are expressed in the form of recursion formula as Equations.\ref{mk}, \ref{sk} and \ref{v}.


\begin{equation}
\label{mk}
{{M}_{i}}=\frac{\text{i}-1}{\text{i}}{{M}_{k-1}}+\frac{1}{\text{i}}{{x}_{\text{i}}}
\end{equation}

\begin{equation}
\label{sk}
{{S}_{i}}={{S}_{i-1}}+({{x}_{i}}-{{M}_{i}})({{x}_{i}}-{{M}_{i-1}})
\end{equation}

\begin{equation}
\label{v}
V=\frac{{{S}_{i}}}{i-1}
\end{equation}

where $i$ is the number of total SNR values, $x_i$ is the latest arrived SNR value, $M_{i-1}$ is latest mean and $\frac{S_{i-1}}{i-2}$ is latest variance. $M_i$ and $V$ are updated mean and variance of all historical SNR values.

A brief demonstration of Equation \ref{mk} is given in Equation \ref{mk proof 1},\ref{mk proof 2} and \ref{mk proof 3}. In addition, a brief demonstration of Equation\ref{sk} is given in Equation \ref{S proof}.

\begin{equation}
\label{mk proof 1}
{{M}_{i}}=\frac{{{x}_{i}}+{{x}_{i-1}}......+{{x}_{1}}}{i}
\end{equation}

\begin{equation}
\label{mk proof 2}
{{M}_{i\text{-}1}}=\frac{{{x}_{i-1}}......+{{x}_{1}}}{k\text{-}1}
\end{equation}

\begin{equation}
\label{mk proof 3}
k{{M}_{i}}-(i-1){{M}_{i\text{-}1}}={{x}_{n}}
\end{equation}

\begin{align}
\label{S proof}
 {{S}_{i}}  & =    \sum\limits_{j=1}^{i}{{{({{x}_{j}}-{{M}_{n}})}^{2}}} \notag \\ 
 & =\sum\limits_{j=1}^{i-1} (x_j-M_{i-1} - \frac{x_i-M_{i-1}}{i})^2+(\frac{(i-1)(x_i-M_{i-1})}{i})^2  \notag \\ 
 & =\sum\limits_{j=1}^{i-1}{{{({{x}_{j}}-{{M}_{i-1}})}^{2}}+\left[ \frac{i-1}{{{i}^{2}}}+\frac{{{(i-1)}^{2}}}{{{i}^{2}}} \right]}{{({{x}_{i}}-{{M}_{i-1}})}^{2}} \notag \\ 
 & ={{S}_{i-1}}+(\frac{i-1}{i}){{({{x}_{i}}-{{M}_{i-1}})}^{2}} \notag \\ 
 & ={{S}_{i-1}}+({{x}_{i}}-{{M}_{i-1}})({{x}_{i}}-{{M}_{i}}) \notag \\ 
\end{align}

As mentioned in the listening process that once a node receives a packet with the same ID with its own, it will change to the forwarding process. Nodes in the forwarding process take responsibility to explore the best forwarder for the next hop. In DBCAR, nodes in forwarding process (the source node) will broadcast a handshake packet to determine communicable neighbors. Each neighbor node receives this handshake packet will reply an ACK packet containing ID, residual energy, and current depth. Then the source node will consider current received SNR values, current depth, residual energy and current SNR fluctuation trends and substitute each value into Equation \ref{DBCAR calculation}, thus obtaining the calculating result of each neighbor node. The neighbor node with the highest calculation result will be chosen as the forwarder for the next-hop. Node ID of the next-hop forwarder will be encoded into the packet transmitted by the source node. The primal parameter of Equation \ref{DBCAR calculation} is $s$ and $\Delta d$. We use NNR-based channel estimation algorithm in Section \ref{NNR Channel Estimation Algorithm} to compute $s$. We traverse $M_{SNR}$ and put each SNR vector into the channel estimation algorithm. Then we get a vector of $s$ corresponding to each neighbor of the source node. 

\begin{equation}
\label{S calculation}
SNR(s, g) = s + \alpha \cdot sgn(g) s
\end{equation}

\begin{equation}
\label{DBCAR calculation}
f(SNR(s, g), \Delta d, m, v, E) = \frac{mE}{v} \Delta d \cdot SNR(s, g)
\end{equation}

In addition, we add a coefficient considering channel stability, residual energy and SNR prediction value. $\frac{m}{v}$ indicate channel stability where $m$ and $v$ are the mean and variance of SNR vectors. The neighbor with a higher mean and lower variance will improve the final calculation result. Besides, $E$ is ratio of the residual energy and initial energy $\frac{E_r}{E_{all}}$in percentage. Even a node has a good channel quality but is running out of energy, the node might not be chosen as the best forwarder. It is an efficient strategy to balance energy so as to extend network lifetime. $SNR(s, g)$ is given in Equation \ref{S calculation} where $g$ is the gradient between $t$ and $t-1$ moments and $sgn$ is the sign function. $\alpha \in [0, 0.3]$ is an impact factor of gradient.





\begin{figure*}[ht]
\begin{multicols}{3}
    \includegraphics[width=2.4in,trim=.5cm .5cm .5cm .5cm]{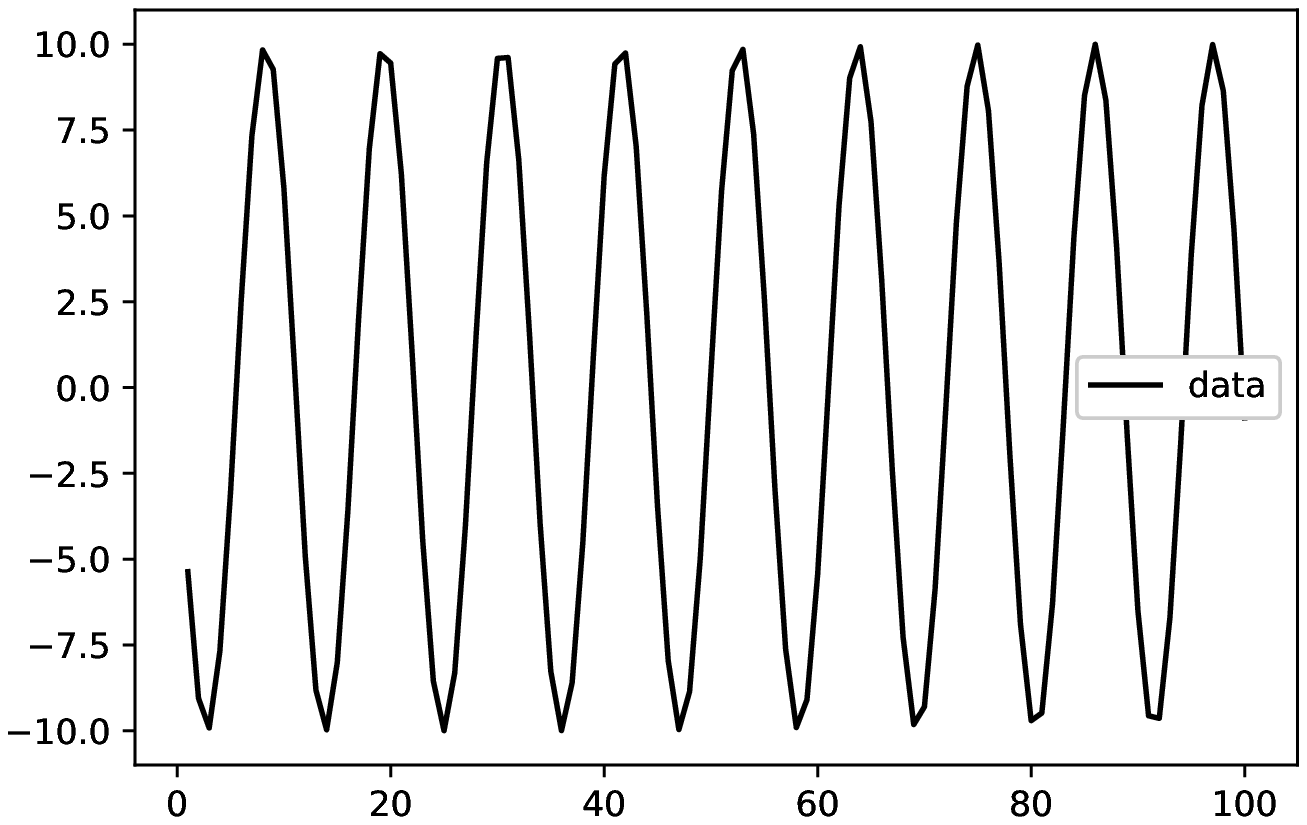}\par\caption{Ideal Period Simulation Samples} \label{ideal period}
    \includegraphics[width=2.4in,trim=.5cm .5cm .5cm .5cm]{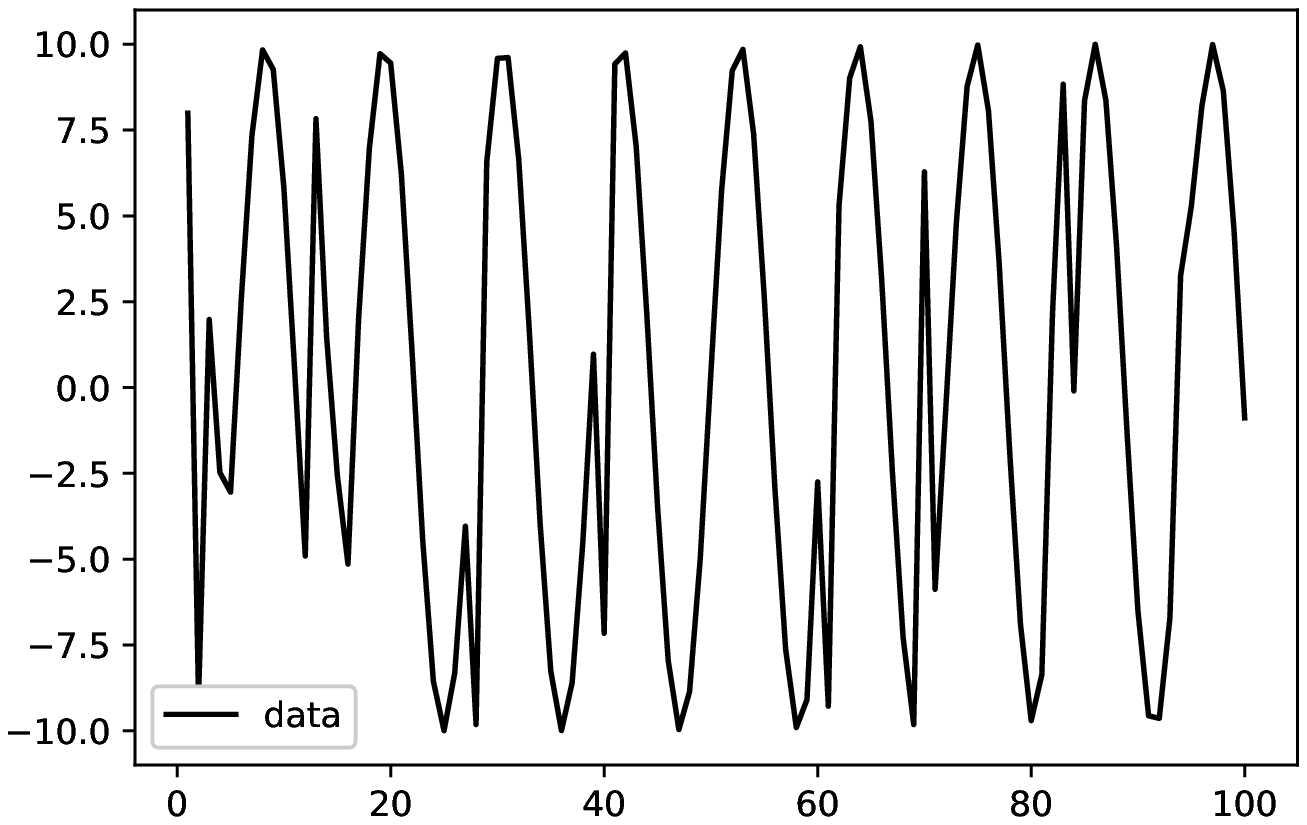}\par\caption{Period with Random Noise Simulation Samples} \label{noise period}
   \includegraphics[width=2.4in,trim=.5cm .5cm .5cm .5cm]{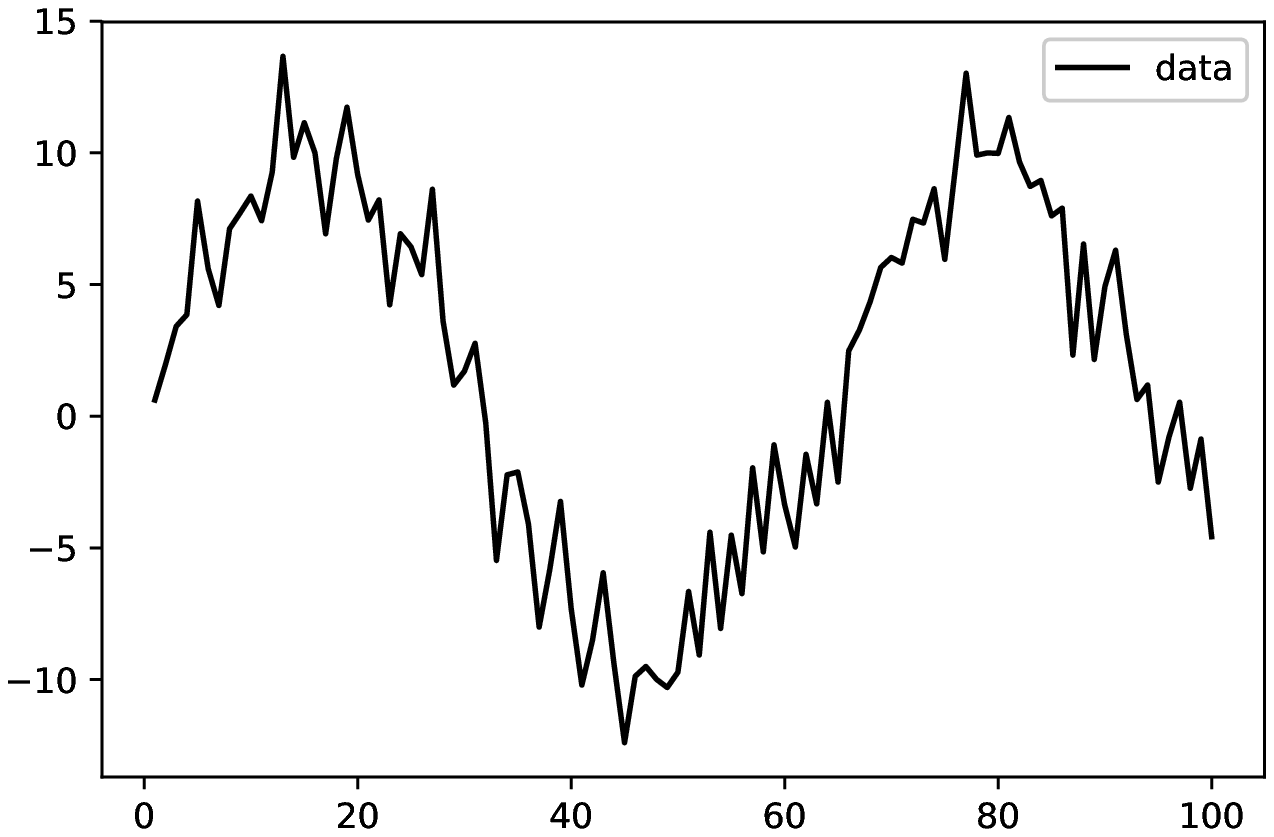}\par\caption{Random Period with Random Noise Simulation Samples} \label{dif noise period}
\end{multicols}
\begin{multicols}{3}
    \includegraphics[width=2.2in,trim=.5cm .5cm .5cm .5cm]{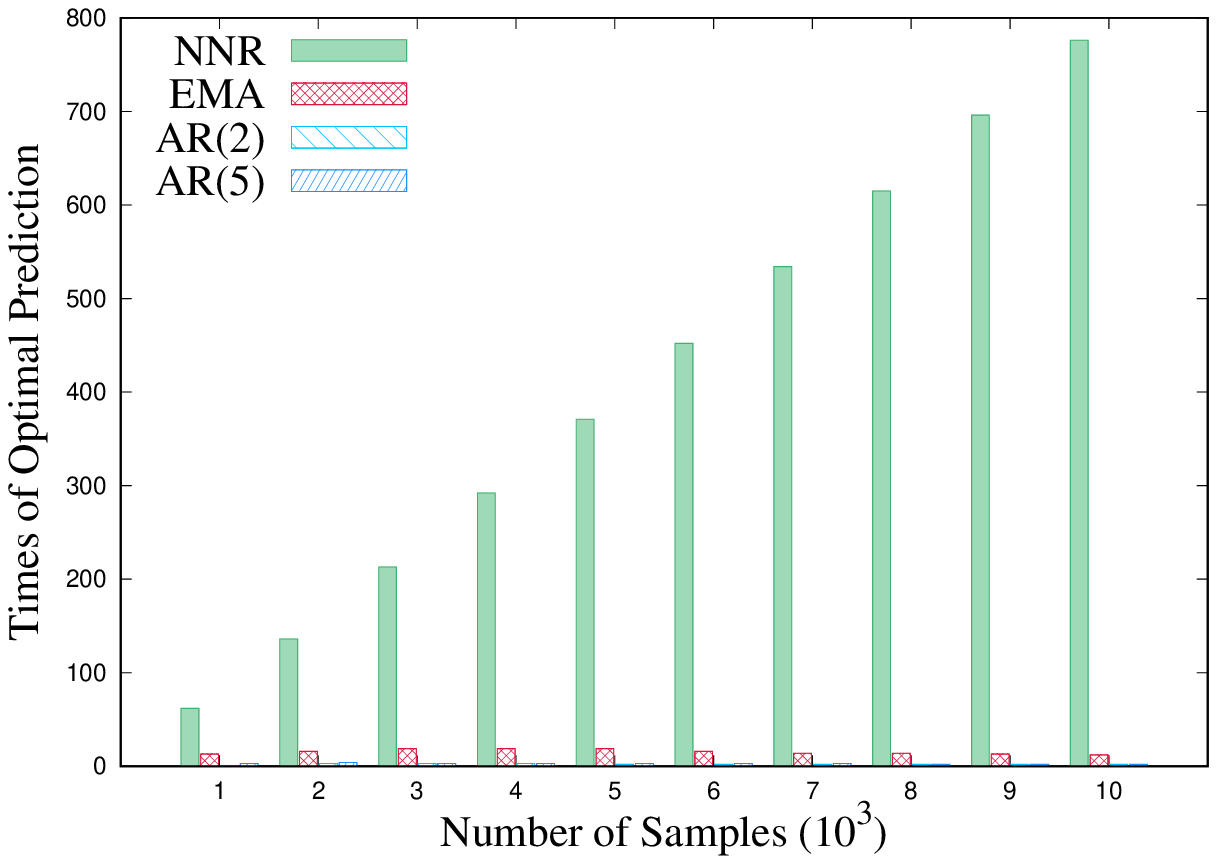}\par\caption{Ideal Period Sample Comparison} \label{ideal period min}
    \includegraphics[width=2.2in,trim=.5cm .5cm .5cm .5cm]{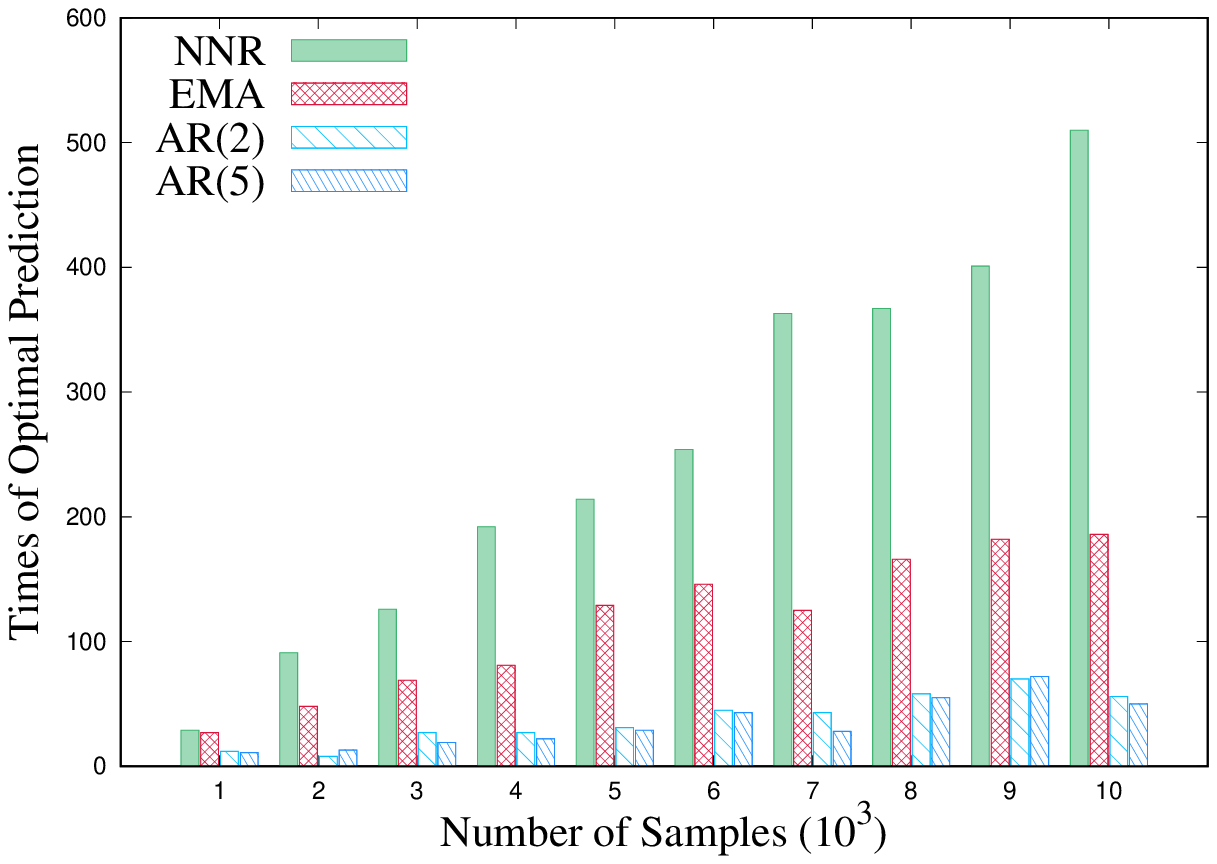}\par\caption{Period with Random Noise Sample Comparison} \label{noise period min}
     \includegraphics[width=2.2in,trim=.5cm .5cm .5cm .5cm]{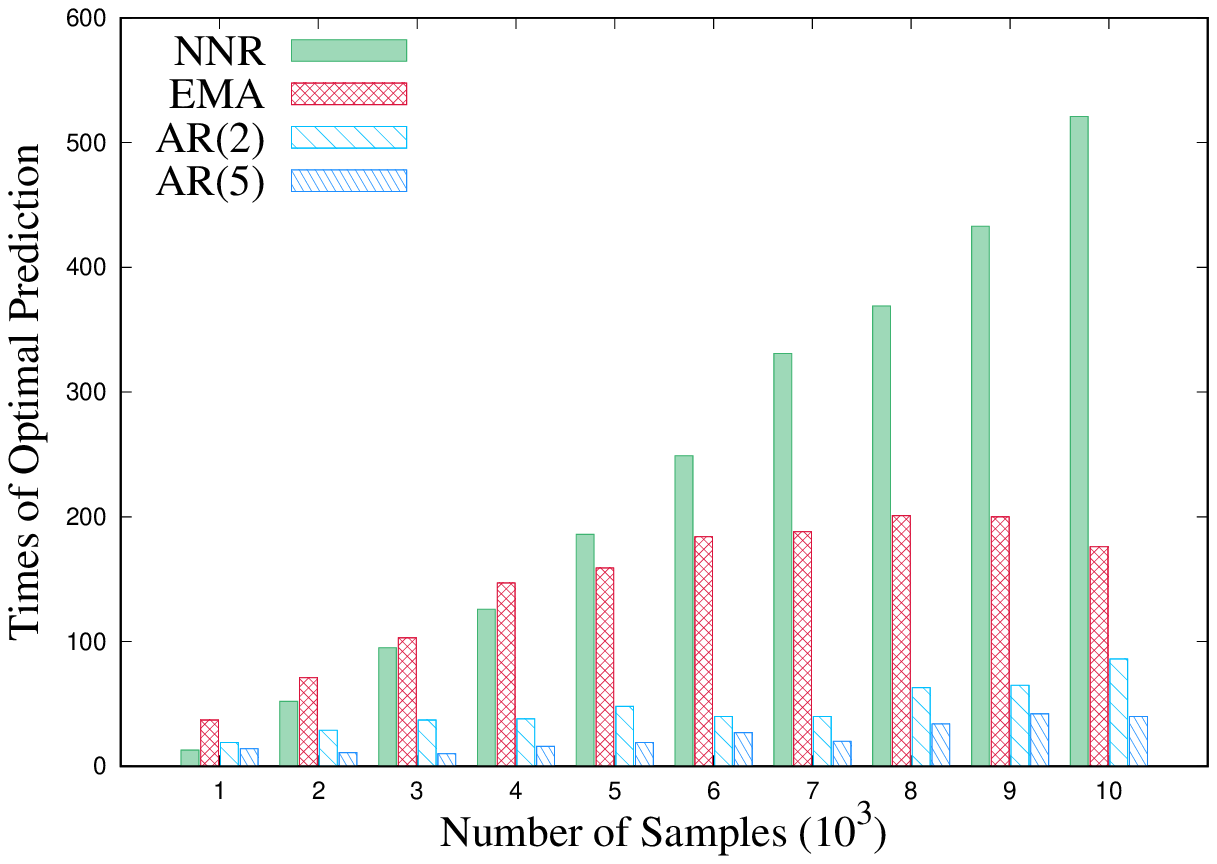}\par\caption{Random Period with Random Noise Sample Comparison}  \label{dif noise period min}
\end{multicols}
\begin{multicols}{3}
    \includegraphics[width=2.2in,trim=.5cm .5cm .5cm .5cm]{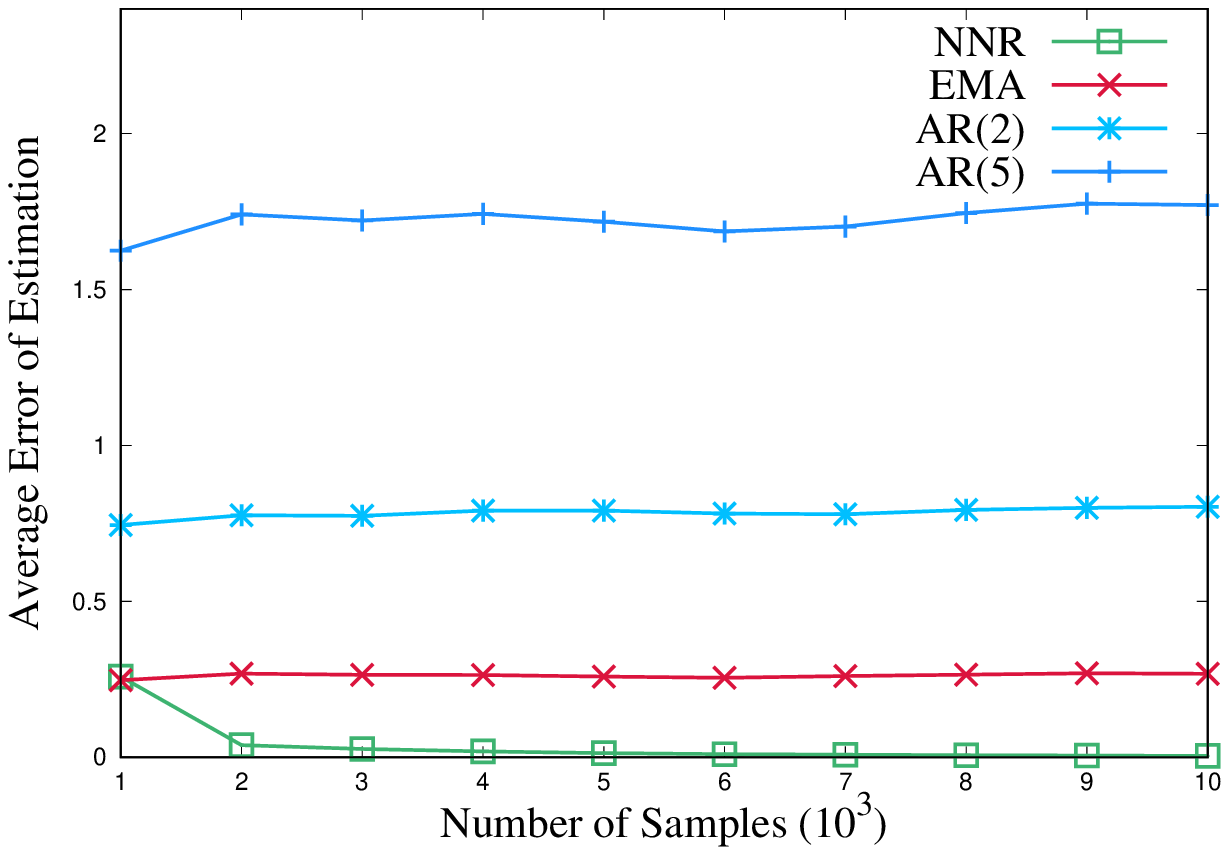}\par\caption{Ideal Period Sample Comparison} \label{ideal period error}
    \includegraphics[width=2.2in,trim=.5cm .5cm .5cm .5cm]{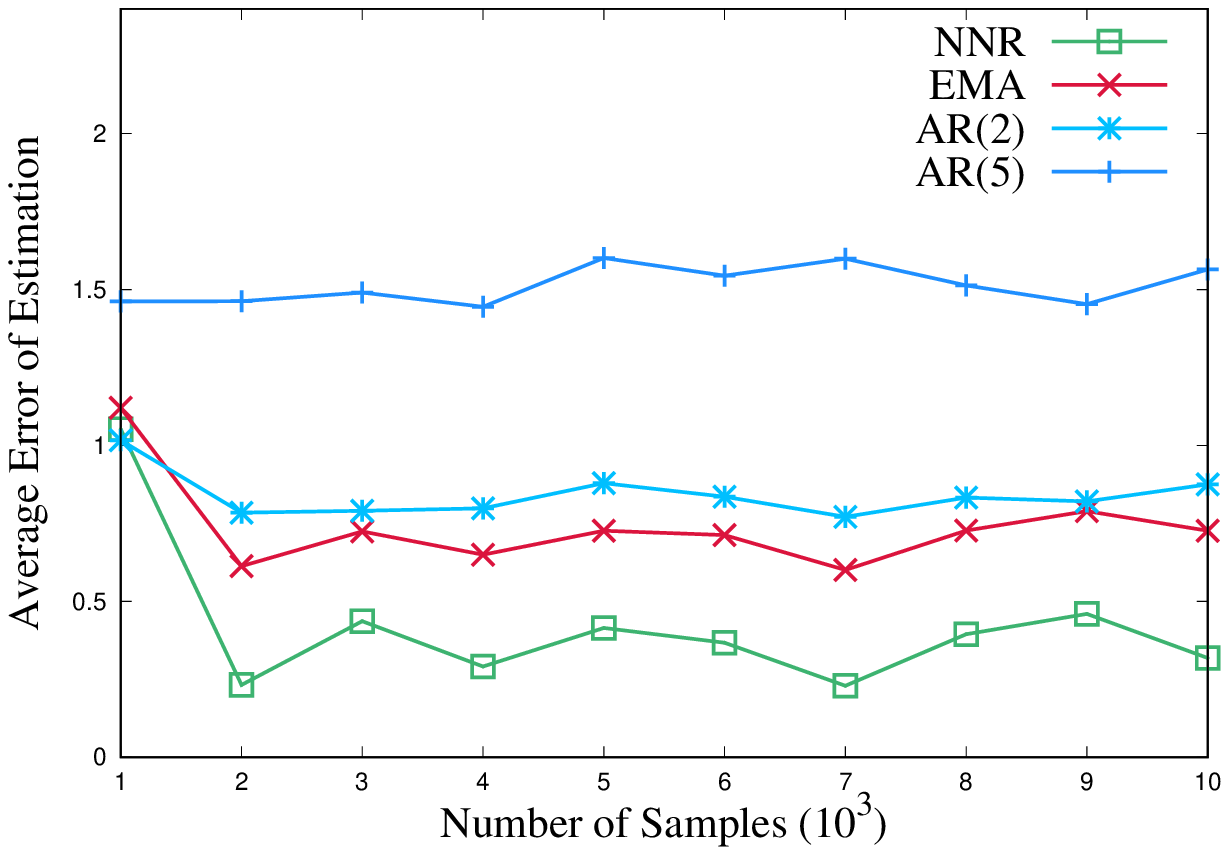}\par\caption{Period with Random Noise Sample Comparison} \label{noise period error}
     \includegraphics[width=2.2in,trim=.5cm .5cm .5cm .5cm]{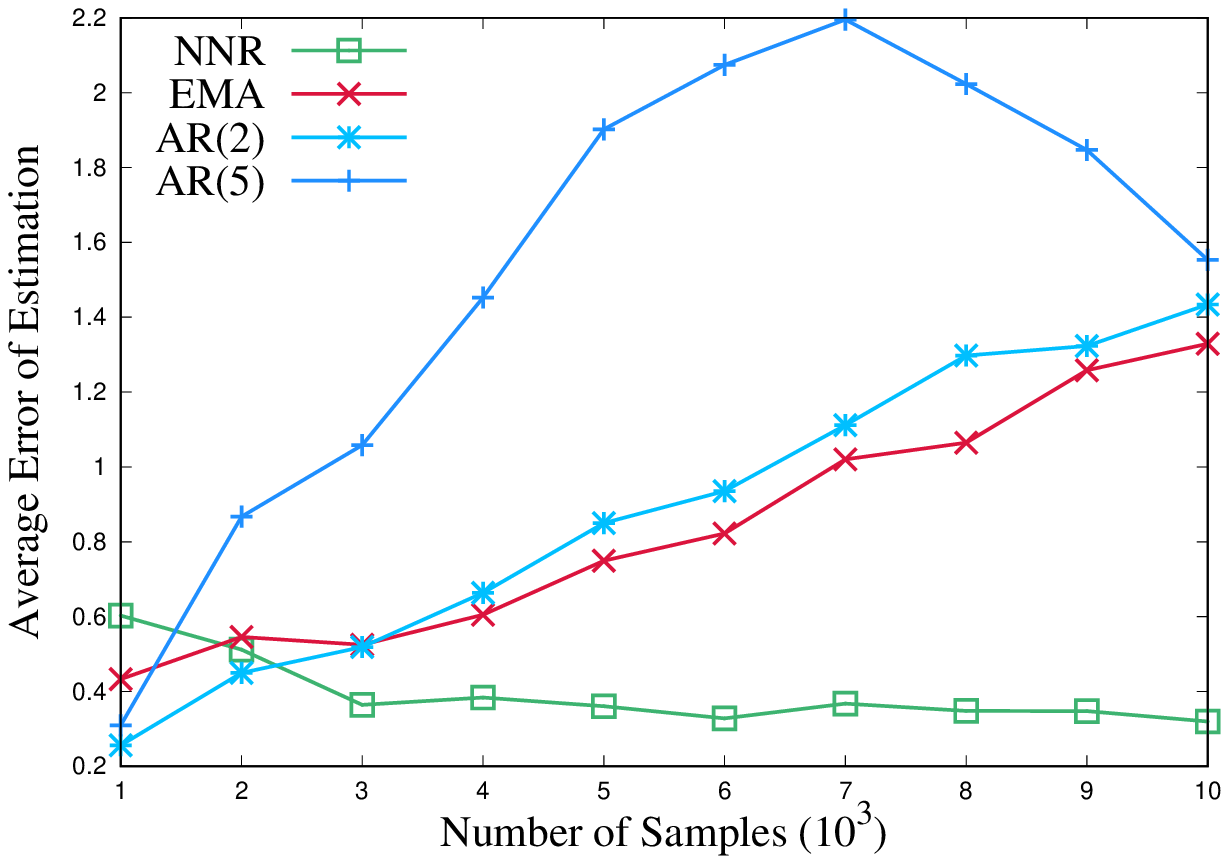}\par\caption{Random Period with Random Noise Sample Comparison}  \label{dif noise period error}
\end{multicols}
\end{figure*}

\section{Performance Evaluations}
\label{simulation}
Performance evaluations consist of three parts: A) Accuracy evaluations of NNR acoustic channel estimation algorithm;  B) Time and space complexity optimization evaluation of NNR acoustic channel estimation algorithm; C) Network performance evaluations of DBCAR with NNR.

\subsection{Accuracy Evaluations of Channel Estimation Algorithms}
\label{channel estimation}

We use both simulation and sea trial data-sets to evaluate proposed method. We simulate ideal period SNR samples, fixed period SNR samples with random noises and random period SNR samples with random noise, receptively. Moreover, we used KAM11 and KW14 sea trial data-sets. For acoustic channel estimation methods, we select EMA, AR(2) and AR(5) to compare with the proposed algorithm. We analyze times of best estimation and prediction errors, demonstrating NNR had an outstanding accuracy than other linear methods.

\begin{figure}[h]
\centering
\includegraphics[width=3in]{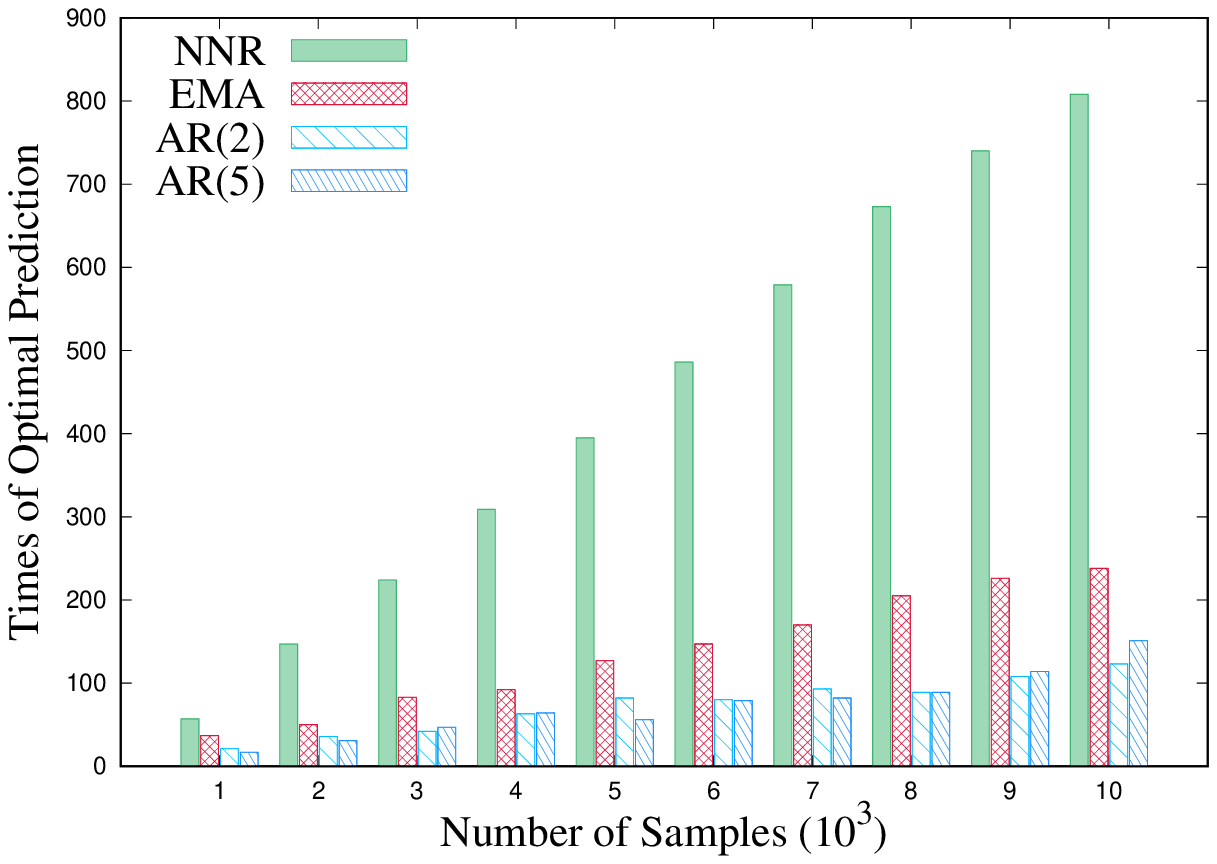}
\caption{Optimal Estimation on KW14 sea trial}
\label{kw14_minrate}
\end{figure}

\begin{figure}[h]
\centering
\includegraphics[width=3in]{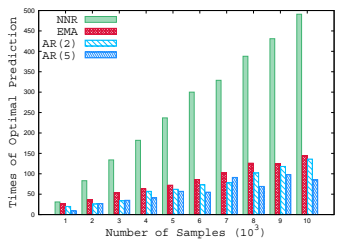}
\caption{Optimal Estimation on KAM11 sea trial}
\label{kam11_minrate}
\end{figure}

\begin{figure}[h]
\centering
\includegraphics[width=3in]{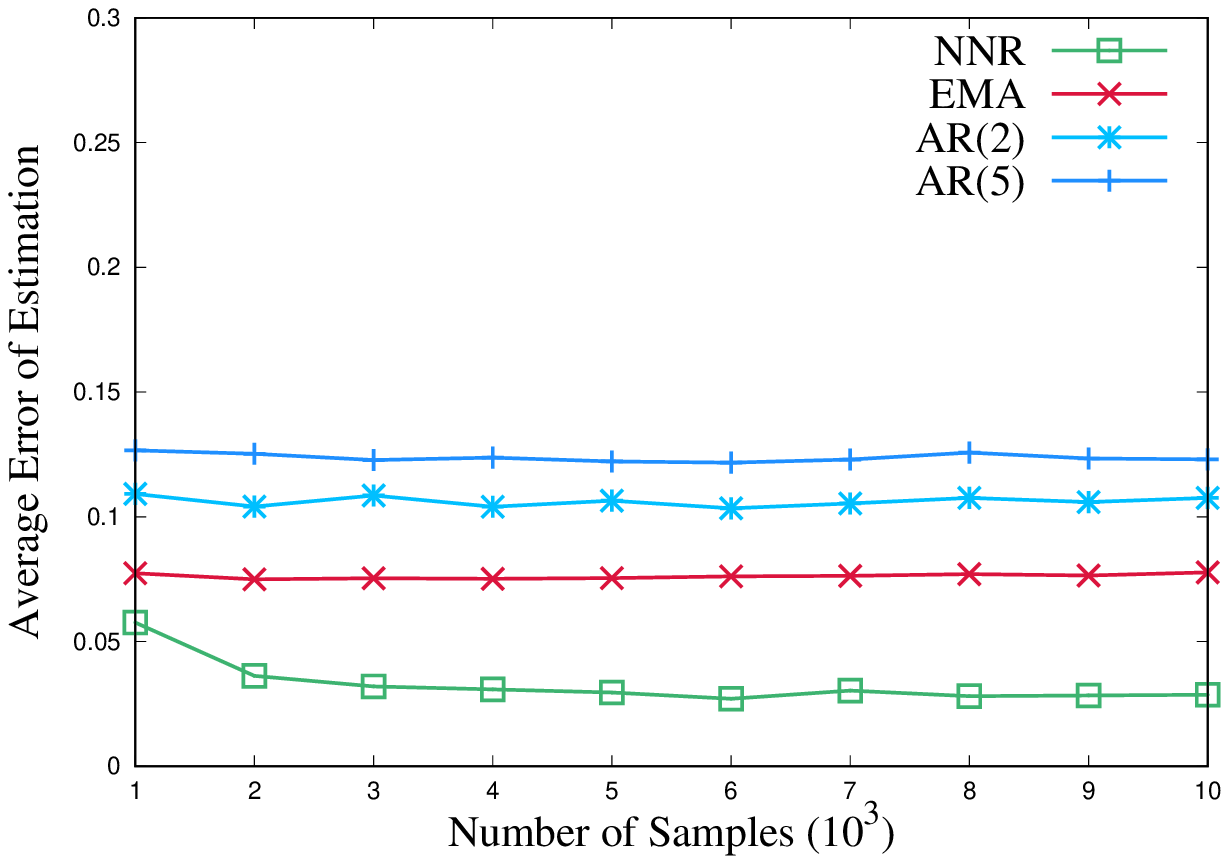}
\caption{Average Estimation Error on KW14 sea trial}
\label{kw14_error}
\end{figure}

\begin{figure}
\centering
\includegraphics[width=3in]{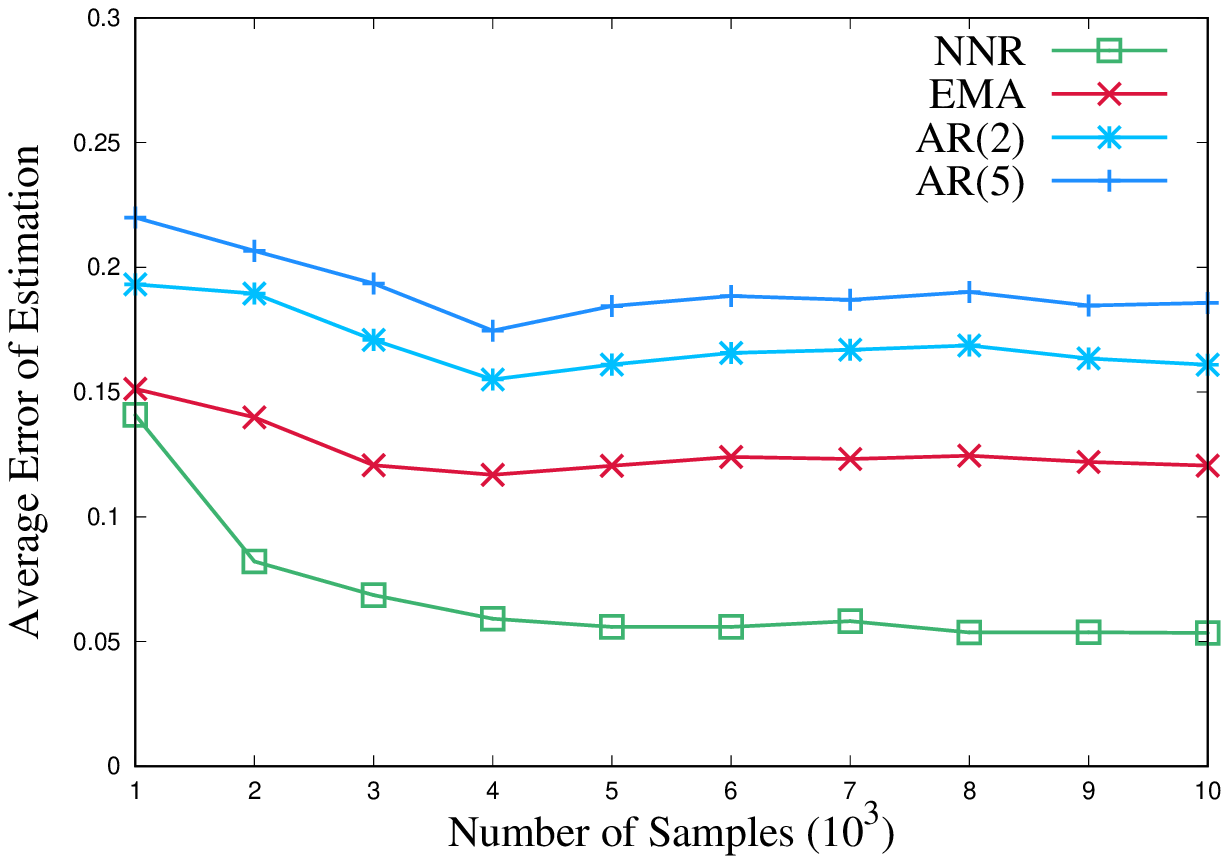}
\caption{Average Estimation Error on KAM11 sea trial}
\label{kam11_error}
\end{figure}

Fig.\ref{ideal period}-\ref{dif noise period} present data trends of ideal period data, period data with random noise and random period data with random noise. Fig.\ref{ideal period min}-\ref{dif noise period min}  show the times of best estimation of the four channel estimation algorithms concerning ideal period data, period data with random noises and random period data with random noises, respectively. The size of the data sets increases from 1000 to 10000. Best estimation is the channel estimation with minimum error in contrary to real SNR values among the four algorithms. In summary, NNR has an average best estimation rate of $83.9\%$ with ideal period samples while the second best is EMA with an average best estimation rate of $4.6\%$, demonstrating that NNR has perfect performance on period samples. However, real received SNR samples are not ideal periodicity. Concerning period data with random noises and random period data with random noises, NNR has an average best estimation rate of $54.2\%$ and $43.4\%$  while the second best is EMA with an average best estimation rate of $22.2\%$ and $31.5\%$. With the size of data sets increasing, the best estimation rate of NNR gradually rises. Fig.\ref{ideal period error}-\ref{dif noise period error} show average channel estimation error of different algorithms on different data-sets. Also, NNR is the best among these channel estimation algorithms. Fig.\ref{kw14}-\ref{kam11} show data trends of sea trials, KAM11 and KW14. From Fig.\ref{kw14_minrate}-\ref{kam11_minrate} together, we may conclude that NNR has an average best estimation rate of $47.8\%$ with KW14 and $42.2\%$ with KAM11 while the second best is MA with an average best estimation rate of $15.8\%$ with KW14 and $15.2\%$ with KAM11. Fig.\ref{kw14_error}-\ref{kam11_error} show average channel estimation error of different methods on sea trial data-sets. We may conclude that NNR is the best among these channel estimation algorithms.

\subsection{Complexity Optimization Evaluations}

In this section, we evaluate time and space complexity optimization in Section \ref{optimization}

\begin{figure}
\centering
\includegraphics[width=3in]{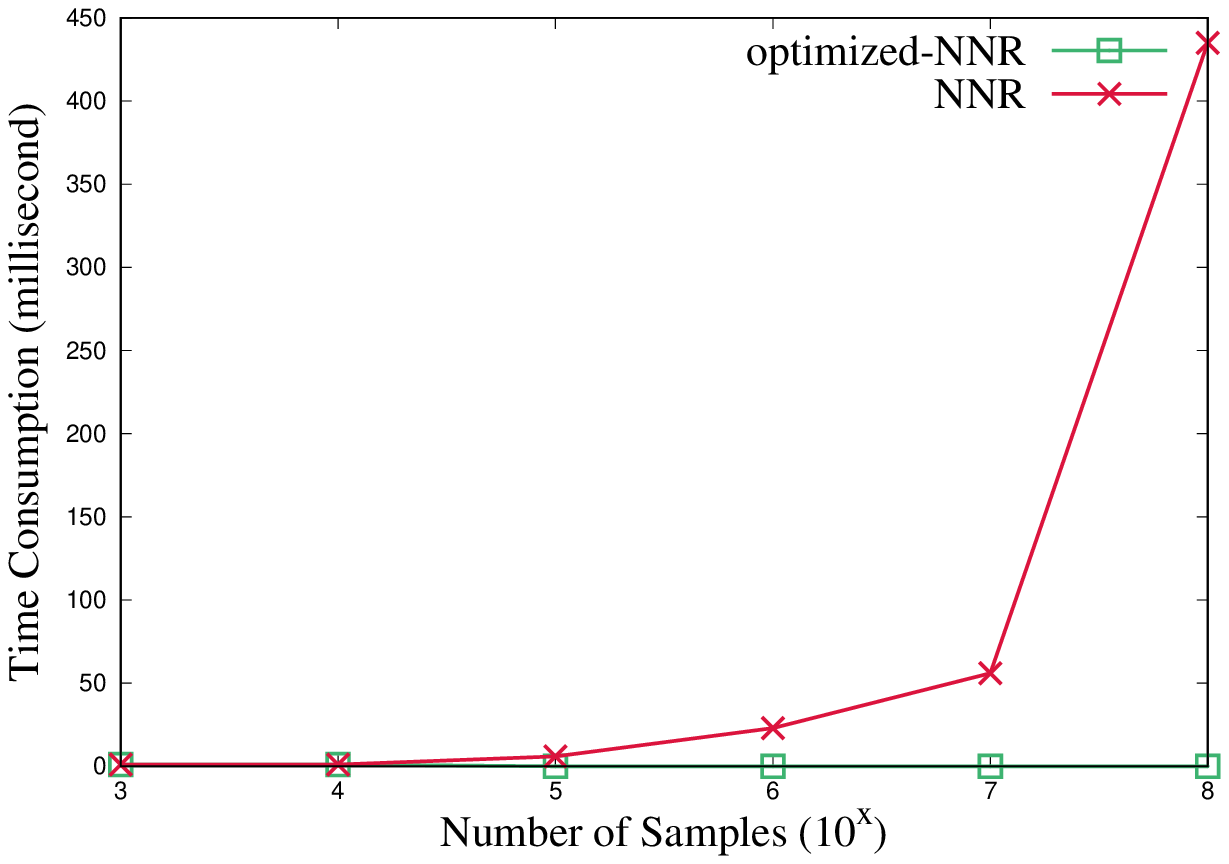}
\caption{Time Complexity Optimization Evaluation}\label{time_complexity}
\end{figure}

\begin{figure}
\centering
\includegraphics[width=3in]{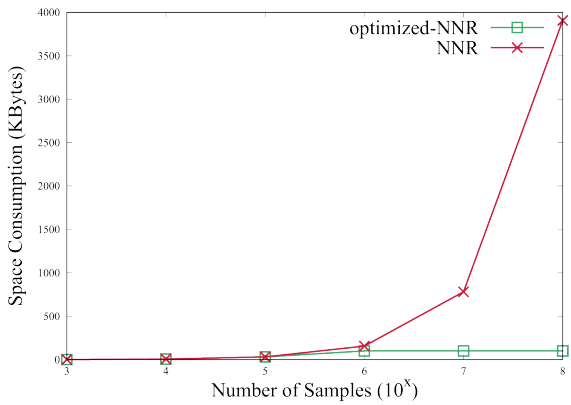}
\caption{Space Complexity Optimization Evaluation}
\label{space_complexity}
\end{figure}

\begin{figure}
\centering
\includegraphics[width=3in]{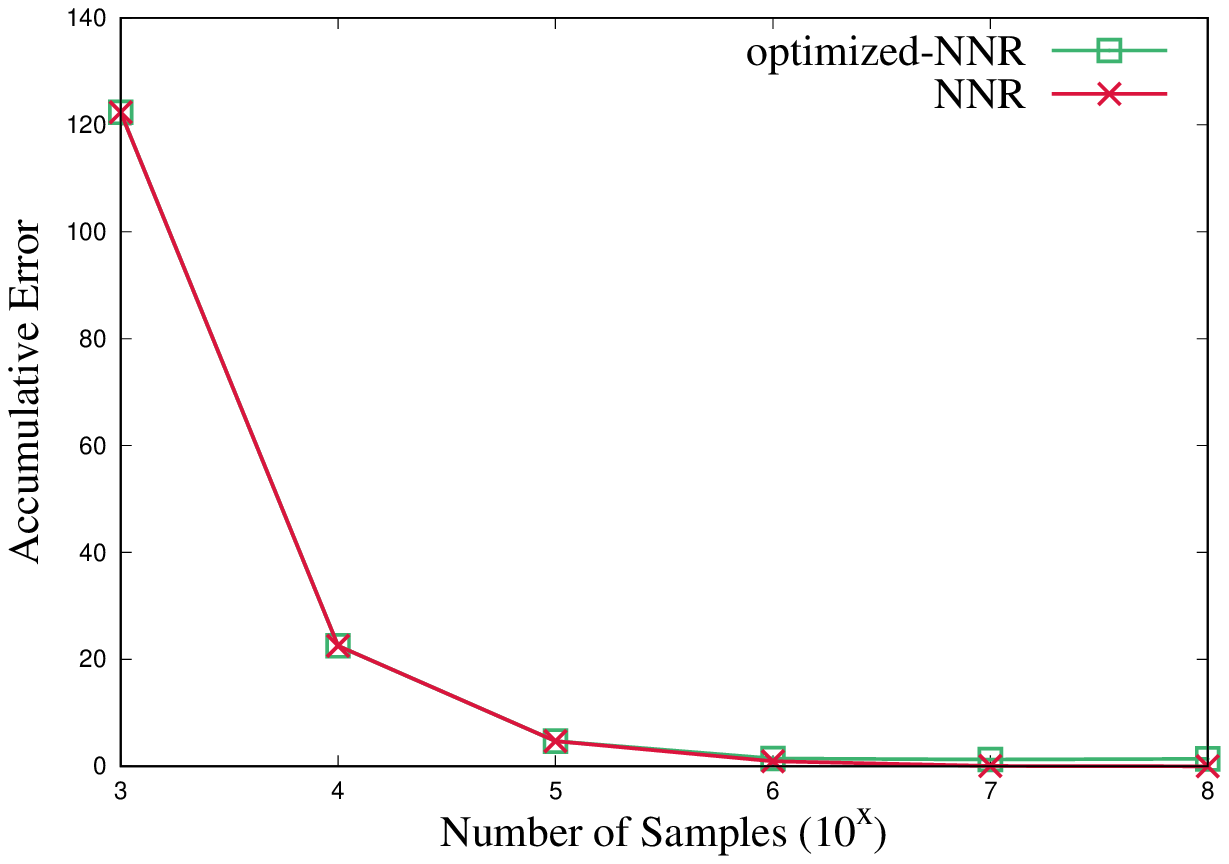}
\caption{Accuracy Evaluation}
\label{accuracy_complexity}
\end{figure}

In Fig.\ref{time_complexity} and Fig.\ref{space_complexity}, we compare time and space efficiency of naive NNR and optimized-NNR. Without loss of generality, we set $k$ to be 3 and $\alpha$, then compare the searching time of $k$ nearest neighbors with the sample size from $10^2$ to $10^8$. Obviously, searching time of optimized-NNR algorithm slightly rises in linear trend while naive NNR in an exponential trend. In addition, we set $L$ to be $10^4$ and $\alpha$ to be $0.2$ in Equation \ref{compression function}. The result is similar to time complexity comparison. Space consumption of optimized NNR algorithm stopped increasing at $L$ while naive NNR rises in an exponential trend. Optimized NNR has an approximate linear complexity in time and space. However, time and space optimization results in tiny errors. Fortunately, estimation accuracy of optimized NNR is very close to naive NNR. The estimation error of optimized NNR is less than $1.48\%$ compared with naive NNR in Fig.\ref{accuracy_complexity}.

\subsection{Network Performance Evaluations}


Network simulations are implemented through Aqua-Sim \cite{aqua2009xie}, a widely used UWSN simulation extension package based on NS-2 \cite{ns2_Varadhan03thens}. Major underwater communication parameters are based on a real-world acoustic modem UWM1000 \cite{LINKQUEST}: bandwidth was $10$ $kbps$; power consumption of sending is $2$ $watt$, receiving is $0.1$ $watt$ and idling is $1\times 10^{-2}$ $watt$. Broadcast MAC protocol is the same as \cite{dbr2008}. $\delta$ in delay function is set to be $\frac{R}{4}$ in the whole simulation. Sensor nodes are randomly deployed in a $500 \times 500 \times 500$ $m^3$ underwater area. Three sinks are uniformly deployed on the surface. The sink nodes are fixed once deployed. For simplicity and without loss of generality, one of the underwater sensor nodes is set to be the source node. In addition, depth of the source node is set to be the max depth and moving slightly and randomly on this plane. Maximal transmission distance $R$ is $150\ m$. Length of data packets is 100 bytes and 20 bytes for handshake packets in DBCAR and CARP. We run the simulation for $10^4$ seconds then statistic major network parameters. The number of nodes increases from 100 to 200 with an interval of 10. The source has a packet rate of $0.1\ packet/s$. Bellhop computes time variable distribution of transmission loss for different locations. Environment data refer to an area in Dickins Seamount located at $54^{\circ} 30' 00'' N, 137^{\circ} 00' 00'' W$. Sound Speed profiles (SSP) and bathymetry profiles are provided in Acoustic Toolbox\cite{toolbox}.

In network performance evaluation, we concern three performance metrics: packet delivery ratio, average end-to-end delay, and average energy cost defined below.

\textit{Packet Delivery Ratio :} the ratio of the number of different packets received at the sink to the number of the total packets transmitted by the source node. This parameter indicates the reliability of a routing protocol.

\textit{Average End-to-End Delay :} the average transmission duration for each packet from the source to the sink. We don't consider lost packets in this work. This parameter indicates time efficiency of a routing protocol.

\textit{Average Energy Consumption :} the average total consumption for each packet received by the sink. Average energy consumption is the ratio of total energy consumption of the network to the number of packets received by the sink. This parameter indicates energy efficiency of a routing protocol.

\begin{figure}[H]
\centering
\includegraphics[width=3in]{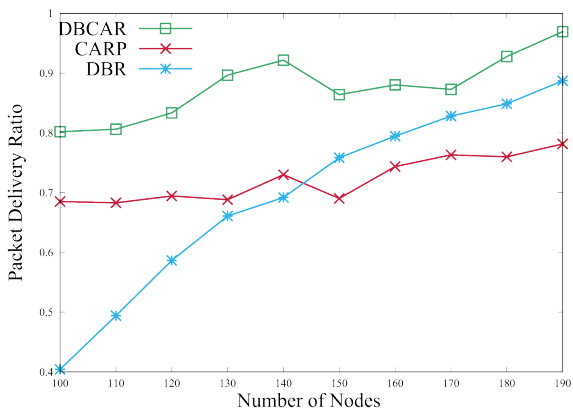}\caption{Packet Delivery Ratio Evaluation}\label{pdr}
\end{figure}

Fig.\ref{pdr} shows packet delivery ratio of DBCAR, CARP, and DBR in terms of different network density. As DBCAR using NNR-based channel quality estimation methods which had an outstanding performance than EMA validated in Section \ref{channel estimation}, DBCAR exceeds DBR without channel quality estimation and CARP with EMA-based channel quality estimation in terms of packet delivery ratio. In contrast with DBR, DBCAR increases packet delivery ratio by 18.2\% on average and 39.7\% in max. In contrast with CARP, DBCAR increases packet delivery ratio by 15.5\% on average and 20.8\% in max.

\begin{figure}[H]
\centering
\includegraphics[width=3in]{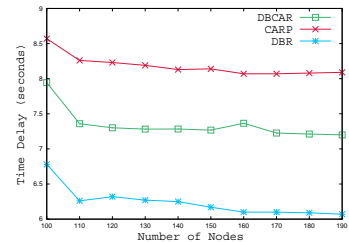}\caption{Average End-to-End Delay Evaluation}\label{time_delay}
\end{figure}

\begin{figure}[H]
\centering
\includegraphics[width=3in]{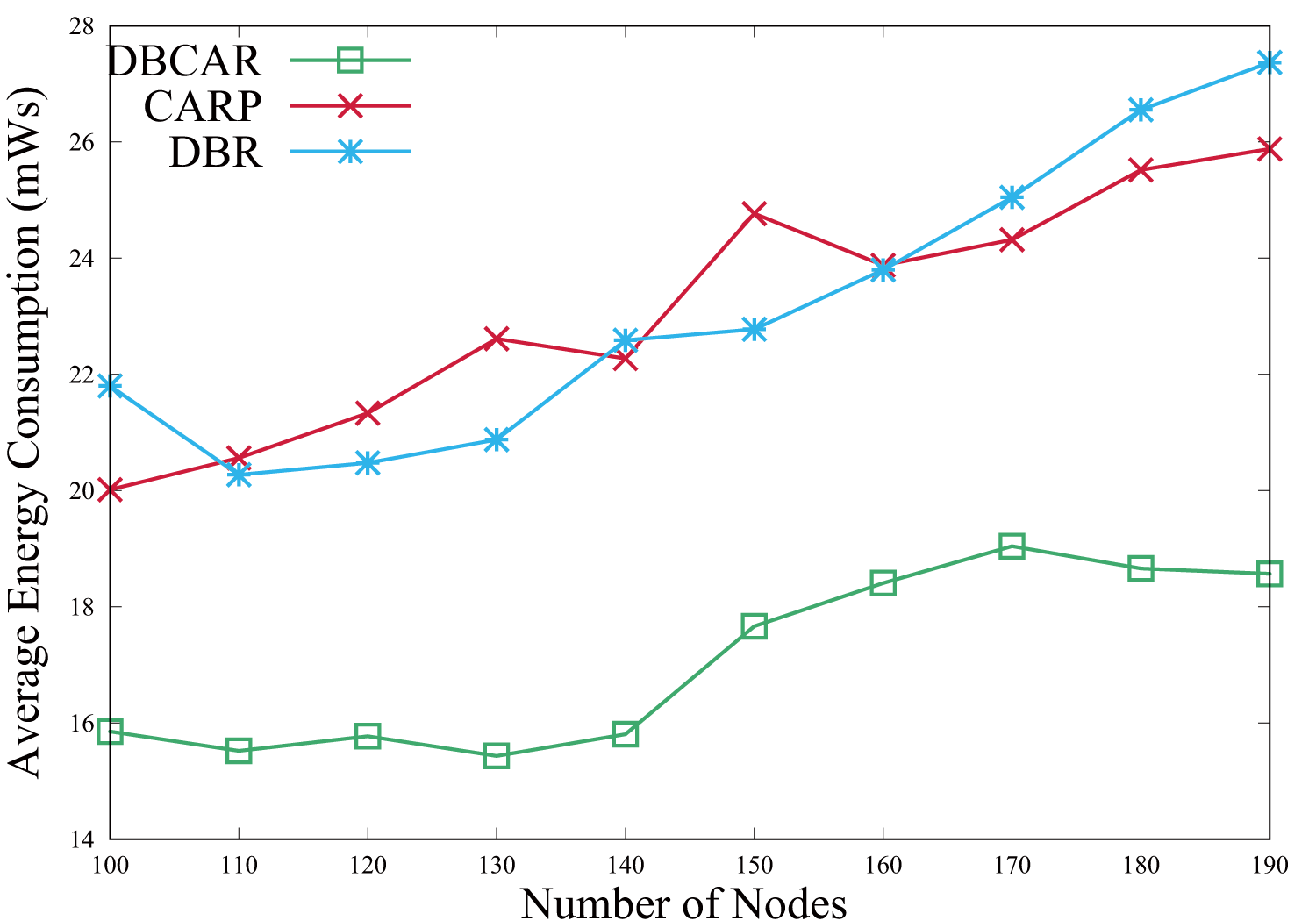}\caption{Average Energy Consumption Evaluation}\label{energy}
\end{figure}

Comparison of average transmission delays of DBCAR, CARP, and DBR is illustrated in Fig.\ref{time_delay}. In Fig.\ref{time_delay}, DBR has a better time efficiency than DBCAR and CARP because DBR is a depth-greedy routing protocol considering no channel parameters. It can explore the neighbor with max depth difference, then make this neighbor as next forwarder without considering channel quality between senders and receivers. As defined above, we only statistics successfully delivered packets in average end-to-end delay results. Although DBR has a lower packet delivery ratio, packets successfully delivered have a shorter transmission delay. Also, DBCAR exceeds CARP in time efficiency. According to simulation results, DBCAR reduces average end-to-end transmission delay by 10.3\% on average and 11.3\% in max in contrast with CARP.

Comparison of Average energy consumption of the three protocols is illustrated in Fig.\ref{energy}. DBCAR has an outstanding energy efficiency than DBR and CARP. Both DBCAR and CARP are single-path routing protocols with a centralized handshake process to select the best forwarder while DBR is a multi-path routing protocol producing inevitable redundancy energy consumption. However, depth-greedy routing reduces hop-counts from sources to sinks for DBR. Therefore, there are little differences in average energy consumption between DBR and CARP. DBCAR took advantages of depth-greedy routing and channel-aware routing, leading to a better energy efficiency. In contrast with DBR, DBCAR reduces average energy consumption by 26.1\% on average and 32.1\% in max. In comparison with CARP, DBCAR reduces average energy consumption by 26.2\% on average and 34.4\% in max.


\section{Conclusions}
\label{conclusion}
In this paper, we introduce NNR for underwater channel estimation. We extend NNR for time series analysis and obtain a better prediction precise for intricately fluctuating and periodical SNR time series. Also, we use a hash table and statistical storage compression to optimize the time and space complexity of NNR quality estimation algorithm. In contrast with linear methods, this algorithm significantly improves prediction accuracy on both simulation and sea trial datasets. With this channel estimation algorithm, we then propose a depth-based channel-aware routing protocol, DBCAR. Taking advantage of depth-greedy forwarding and channel-aware reliable communication, DBCAR has an outstanding network performance on packet delivery ratio, average energy consumption and average transmission delay which is validated through extensive simulations.


\ifCLASSOPTIONcaptionsoff
  \newpage
\fi

\bibliographystyle{ieeetr}
\bibliography{uwsnbib}

\begin{IEEEbiography}[{\includegraphics[width=1in,height=1.25in,clip,keepaspectratio]{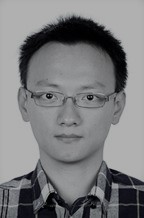}}]{Boyu Diao}
received his B.Eng. degree in computer science from Beijing Institute of Technology, Beijing, China, in 2012 and became He a Ph. D. candidate of Institute of Computing Technology, Chinese Academy of Sciences, Beijing, China, in 2015. He is also a Ph. D. candidate of University of Chinese Academy of Sciences, Beijing, China. His research focuses on underwater sensor networks, distributed algorithms and  multi-sensor data fusion.
\end{IEEEbiography}

\begin{IEEEbiography}[{\includegraphics[width=1in,height=1.25in,clip,keepaspectratio]{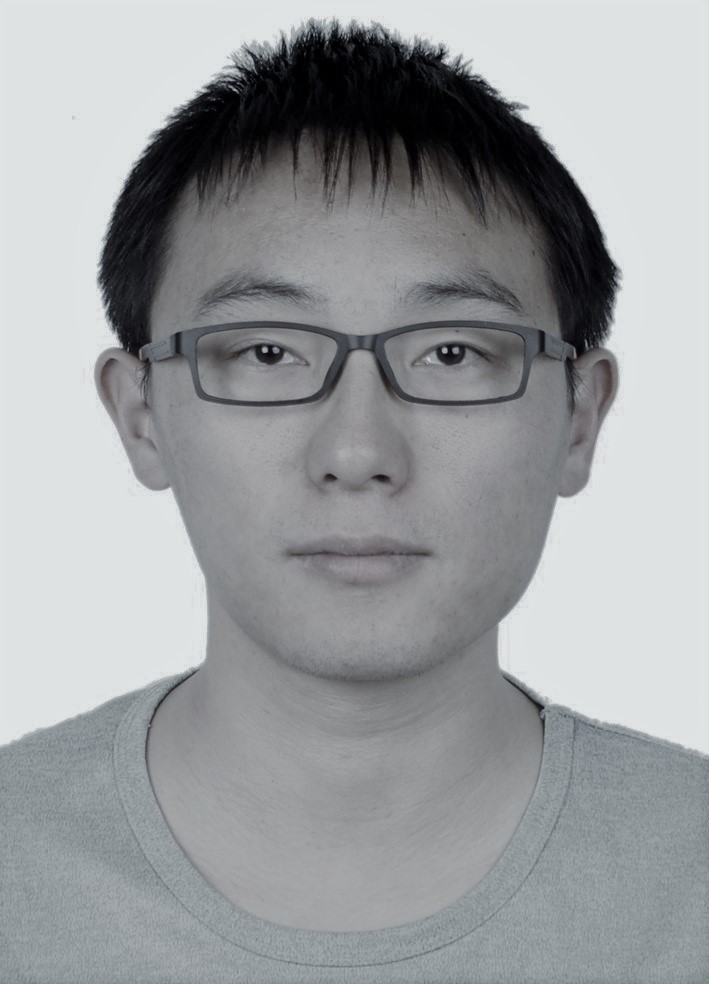}}]{Chao Li}
received B.Eng. degree from the Southwest University, Chongqing, China, in 2008, received his M.Eng. degree in computer science from Beijing Normal of Technology, Beijing, China, in 2011. He received his PHD degree in 2016 from Institute of Computing Technology (ICT), Chinese Academy of Sciences.
His research interest includes the wireless sensor networks, the underwater sensor networks and distributed algorithms, and new trends in cloud platform.

\end{IEEEbiography}

\begin{IEEEbiography}[{\includegraphics[width=1in,height=1.25in,clip,keepaspectratio]{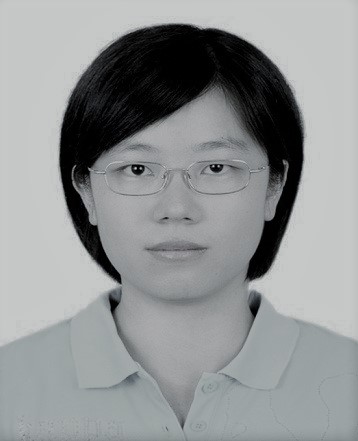}}]{Qi Wang}
is an assistant professor at Institute of Computing Technology, Chinese Academy of Sciences (ICT/CAS) in Beijing, China. She received the Ph.D degree in computer science from Chinese Academy of Sciences, Beijing, China in 2015. 
Her research focuses on performance evaluation and optimization for wireless ad hoc and sensor networks.
\end{IEEEbiography}

\begin{IEEEbiography}[{\includegraphics[width=1in,height=1.25in,clip,keepaspectratio]{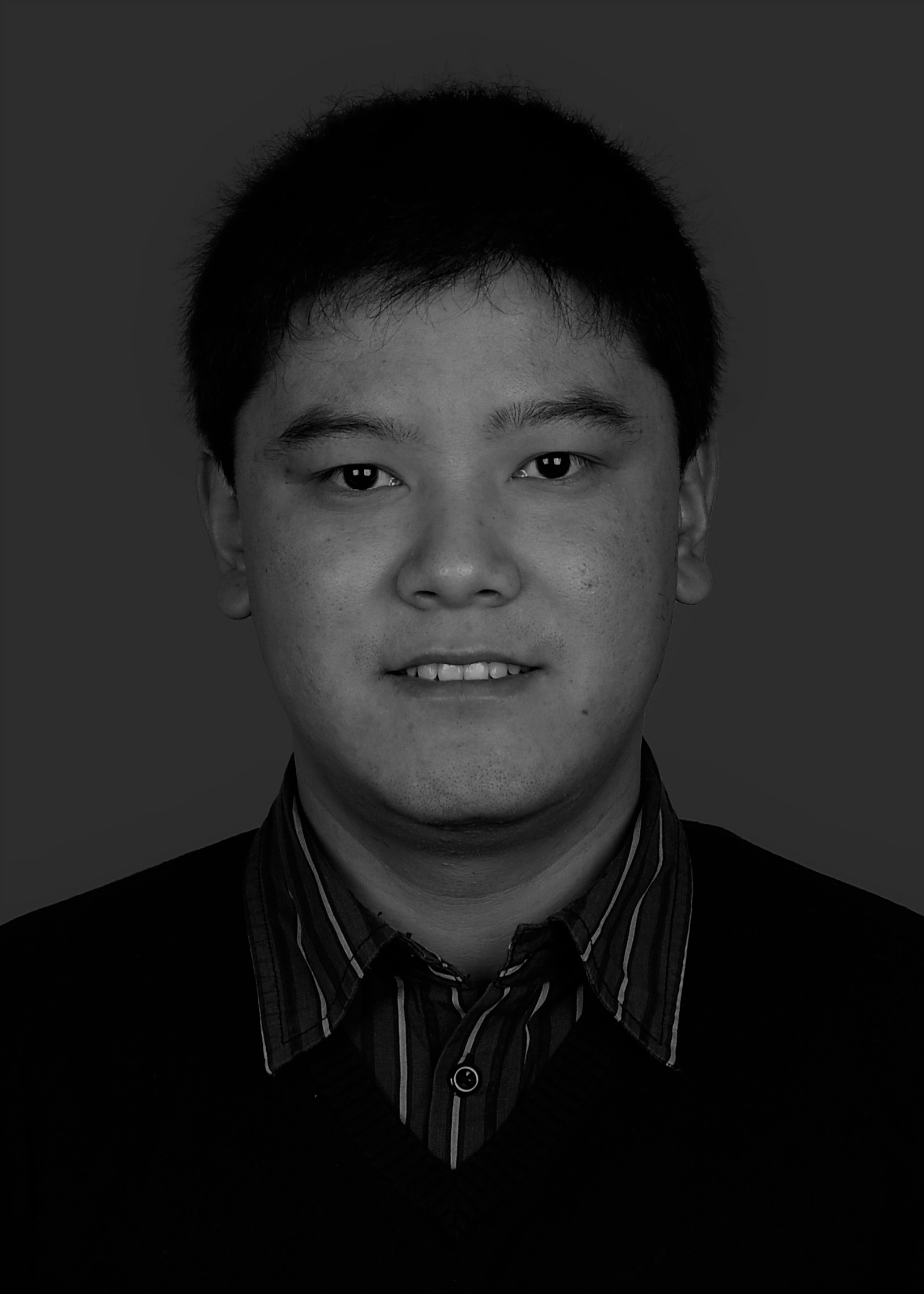}}]{Zhulin An}
received the B.Eng. and M.Eng. degrees in Computer Science from School of Computer and Information, Hefei University of Technology, Hefei, China, in 2003 and 2006, respectively. He received his PHD degree in 2009 from Institute of Computing Technology (ICT), Chinese Academy of Sciences and became an associate professor in 2014.
His research interests include parallel and distributed algorithms and time synchronization in wireless sensor networks.
\end{IEEEbiography}

\begin{IEEEbiography}[{\includegraphics[width=1in,height=1.25in,clip,keepaspectratio]{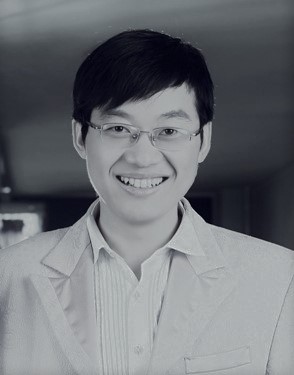}}]{Yongjun Xu}
(M'06) received his B.Eng. degree in computer communication from Xi'an Institute of Posts \& Telecoms (China) in 2001, and then entered Institute of Computing Technology (ICT), Chinese Academy of Sciences as a graduate student. He received his PHD degree in 2006 and became an associate professor in 2008. He is now a professor in Institute of Computing Technology, Chinese Academy of Sciences.
His current research interests include cyber-physical systems, multi-sensor data fusion.
\end{IEEEbiography}

\EOD

\end{document}